\documentclass[letterpaper,twocolumn,english,aps,pra,showpacs,preprintnumbers]{revtex4}
\usepackage{lmodern}
\usepackage[T1]{fontenc}
\usepackage[latin9]{inputenc}
\usepackage{units}
\usepackage{amsmath}
\usepackage{amssymb}
\usepackage{graphicx}
\usepackage{esint}

\makeatletter

\@ifundefined{textcolor}{}
{%
 \definecolor{BLACK}{gray}{0}
 \definecolor{WHITE}{gray}{1}
 \definecolor{RED}{rgb}{1,0,0}
 \definecolor{GREEN}{rgb}{0,1,0}
 \definecolor{BLUE}{rgb}{0,0,1}
 \definecolor{CYAN}{cmyk}{1,0,0,0}
 \definecolor{MAGENTA}{cmyk}{0,1,0,0}
 \definecolor{YELLOW}{cmyk}{0,0,1,0}
 }

\makeatother

\usepackage{babel}
\begin{document}

\title{Superradiance in spin-$J$ particles: Effects of multiple levels}

\author{G.-D. Lin and S. F. Yelin}

\affiliation{Department of Physics, University of Connecticut, Storrs, CT 06269\\
ITAMP, Harvard-Smithsonian Center for Astrophysics, Cambridge, MA
02138}
\begin{abstract}
We study the superradiance dynamics in a dense system of atoms each
of which can be generally a spin-$j$ particle with $j$ an arbitrary
half-integer. We generalize Dicke's superradiance point of view to
multiple-level systems, and compare the results based on a novel approach
we have developed in {[}Yelin \textit{et al.}, arXiv:quant-ph/0509184{]}.
Using this formalism we derive an effective two-body description that
shows cooperative and collective effects for spin-$j$ particles,
taking into account the coherence of transitions between different
atomic levels. We find that the superradiance, which is well-known
as a many-body phenomenon, can also be modified by multiple level
effects. We also discuss the feasibility and propose that our approach
can be applied to polar molecules, for their vibrational states have
multi-level structure which is partially harmonic.
\end{abstract}

\pacs{42.50.Nn, 42.50.Ar, 33.20.Tp}

\maketitle

\section{Introduction\label{sec:Introduction}}

Quantum many-body physics has been one of the most attractive areas
for decades along with the remarkable advents in the fields of ultracold
atomic and molecular systems and quantum optics. These systems not
only provide an excellent testbed to study quantum nature of various
many-body phenomena such as Bose-Einstein condensation, superfluidity
\cite{Sci95_Anderson,Nat03_Greiner,RMP01_Leggett}, quantum magnetism,
and quantum phase transitions \cite{PRB89_Fisher,Nat02_Greiner,Nat06_Hadzibabic,PRL08_Weimer},
but also inspire the implementation of quantum machinery such as quantum
simulation \cite{PRL03_Duan,PRL04_Porras,NatPhys08_Friedenauer,Nat10_Gerritsma,Nat10_Kim,Nat09_Lin}
and quantum computing \cite{PRL99_Brennen,Kuznetsova11a,Kuznetsova11b}.
In many ways, quantum many-body effects are {}``exotic'' compared
to their classical counterparts, and even to quantum single-body physics,
mainly due to particle statistics and indistinguishability of particles.
The circumstances can become even more complicated when an ensemble
of particles interact cooperatively which results in higher-order
nonlinear effects. Superradiance, usually representing an $N^{2}$
enhancement of the radiation intensity due to coherent decay of a
dense sample consisting of $N$ excited atoms, is one important example
which can be understood qualitatively through particle indistinguishability
and symmetry arguments without the need for considering particle statistics.
This phenomenon was first predicted in 1954 by R. H. Dicke \cite{PhysRev54_Dicke},
who pointed out that the radiative properties of an excited atom can
be very different just because other atoms are present or not, given
that their distance is much smaller than the wavelength of the radiation
field even if the particle wavepackets do not overlap and no direct
interaction is present. Since then such cooperative effects have been
intensively investigated both theoretically and experimentally \cite{PhysLett74_DeMartini,PhysRep82_Gross,Sci99_Inouye,PRL99_Moore,PRA00_Mustecaplioglu,PRL03_Farooqi,PRL08_Akkermans}.
Recently, superradiance re-catches one's attention in the context
of a Bose-Einstein condensate coupled to a cavity \cite{Nat10_Baumann,PRL10_Nagy},
alkaline-earth-metal atoms \cite{PRA10_Meiser}, Rydberg atoms \cite{PRL03_Farooqi,PRA07_Wang},
and quantum dots \cite{NatPhys07_Scheibner}, as well as its strong
connections to quantum information through the so-called Dicke states.
Such states are fully symmetric states by particle permutation, and
mostly serve as the main stage during the superradiance process.

Traditionally, superradiance deals with two-level atoms or other spin-$\nicefrac{1}{2}$
systems, being first excited, that decay cooperatively. It is natural
as a next step to consider particles with larger spins, i.e. systems
with a multiple level structure. Examples include the near-harmonic
level structure of low-lying vibrational states in molecules. Generally
speaking, multi-level structure brings up more complications to the
radiating system. For example, an excited atom in a higher level is
still excited after emitting a photon; the atoms and photons from
different-level transitions can further cooperate and modify the overall
emission behavior. In order to study multi-level effects, we re-consider
Dicke's point of view of superradiance as the starting point by first
assuming the system to be point-like and fully symmetric.

However, Dicke's picture is only qualitatively correct and insufficient
to describe real situations, where the actual arrangement of particles,
the sample's finite size, and dipole-dipole interactions play a role.
Microscopically, single atoms build up inter-atomic coherence due
to virtual photon exchange caused by dipole-dipole interaction and
form many-body states such as the Dicke states \cite{PhysLett74_DeMartini,PhysRep82_Gross}.
The coherence can be breached when the geometry of particle arrangement
introduces inhomogeneity such as dipole-dipole interaction between
each pair of particles. This leads to dephasing effects and therefore
Dicke's picture fails to be valid. To characterize how the {}``finite
size'' influences superradiant behavior, a parameter, cooperativity
$\mathcal{C}\sim\mathcal{N}\lambda^{3}$ is introduced, with $\mathcal{N}$
the number density and $\lambda$ the wavelength of the transition
field. One then expects that superradiance is observable for $\mathcal{C}\gg1$
and be suppressed for small $\mathcal{C}$. Our previous study \cite{PRA07_Wang}
has further suggested a more accurate estimation that the criterion
of observing superradiance is approximately given by $\mathcal{N}\lambda^{2}d$
($d$ is the sample size), in agreement with \cite{PRL08_Akkermans},
where such a factor is found to be an essential one that determines
whether cooperative effects dominate or not. To take into account
the realistic arrangement of our particle systems, we use a novel
formalism that considers only two probe particles, treating the spread
of environment atoms in the mean-field approximation and then take
an average over all possible particle pairs \cite{PRA99_Fleischhauer,Yelin05}.
This approach enables us to write down an effective master equation,
retaining the degrees of freedom of two-body coherence which can be
regarded as a projection of the many-body coherence in the original
system. This method has been proven to show a good agreement with
ongoing experiments with Rydberg atoms \cite{PRA07_Wang}. In this
paper we further apply this formalism to spin-$j$ systems. 

This article is organized as follows: In Sec. \ref{sec:Dicke} we
discuss the original picture proposed by Dicke and generalize the
idea to multi-level systems. Sec. \ref{sec:formalism} sketches the
formalism developed in \cite{PRA99_Fleischhauer,Yelin05} and summarize
the governing equations. We then apply this method to multiple levels
and show the results in Sec. \ref{sec:result}. There we also discuss
the differences from the Dicke model and investigate the significance
of many-body and multi-level correlations. In Sec. \ref{sec:Doppler}
we further consider the thermal Doppler broadening and calculate the
marginal conditions that superradiance can tolerate. Finally, in Sec.
\ref{sec:vibration} a dipolar molecular gas is discussed as an example,
for which we consider the vibrational states and investigate the superradiance
effects from its vibrational states.

\section{Dicke superradiance\label{sec:Dicke}}

\begin{figure}
\begin{centering}
\includegraphics[width=6cm]{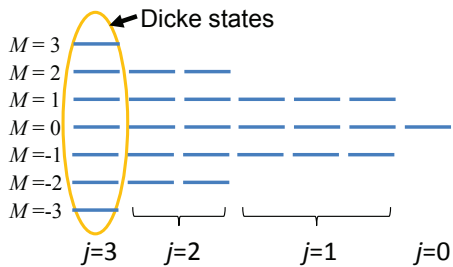}
\par\end{centering}

\caption{\label{fig:level}(color online). The energy level structure for $N=3$
three-level atoms ($j=1$). There are $(2j+1)^{3}=27$ levels, including
the fully symmetric $2Nj+1=7$ states (the Dicke states).}
\end{figure}

To gain qualitative understanding of the Dicke superradiance, we start
with considering an ensemble of ($2j+1$)-level atoms confined within
a small region with its size much smaller than the wavelength of the
radiation field. In this limit, the particles are indistinguishable
viewed by the field and must be regarded as a whole quantum object.
To emphasize that the collective radiative behavior is governed solely
by many-body effects, we do not assume any instantaneous, i.e. non-radiative,
interaction between atoms. The inter-particle spacing is large enough
so that the overlap of particle wavepackets is negligible. In other
words, the exchange interaction plays no role as well as the fermionic
or bosonic nature of the atoms. Suppose that the transitions are induced
by dipoles through the interaction Hamiltonian $V=-\sum_{i}\vec{p}_{i}\cdot\vec{E}(\vec{r}_{i})$,
where $\vec{p}_{i}$ is the dipole operator of the $i$th atom and
$\vec{E}(\vec{r}_{i})$ is the local field at the coordinate $\vec{r}_{i}$.
Under the long-wavelength assumption of the field for a given small
system size, the spatial dependence can be eliminated. Therefore $V=-\vec{E}\cdot\sum_{i}\vec{p}_{i}=-\sum_{\mu=x,y,z}\wp_{\mu}(E_{\mu}^{-}\hat{D}^{-}+\mbox{h.c.})$
in the rotating wave approximation. Here $\wp_{\mu}$ is the dipole
moment magnitude of an atom in the $\mu$ direction, $E_{\mu}^{\pm}$
is the positive (negative) frequency component of the field, and $\hat{D}^{\pm}\equiv\sum_{i}\sigma_{i}^{\pm}$
with $\sigma_{i}^{-}\equiv\sum_{m=-J}^{J-1}|m\rangle_{i}\langle m+1|$
and $\sigma_{i}^{+}\equiv(\sigma_{i}^{-})^{\dagger}$. Note that $V$
does not break the permutational symmetry of the particles. If all
the atoms are initially excited, time evolution will only take the
state of the system around the fully symmetric manifold, whose eigenstates
are usually called the Dicke states (See Fig. \ref{fig:level}, \cite{PhysRev54_Dicke,PhysRep82_Gross}):
\begin{equation}
|J,M\rangle=\sqrt{\frac{(J+M)!}{(2J)!(J-M)!}}(\hat{J}^{-})^{J-M}|J,J\rangle,\label{eq:state_JM}
\end{equation}
where $J=Nj$ is the total spin of $N$ spin-$j$ atoms and the integer
$M$ denoting the level index can only go from $J$ through $-J$;
the total spin ladder operators $\hat{J}^{\pm}=\sum_{i}\hat{j}_{i}^{\pm}$
satisfy $\hat{J}^{\pm}|J,M\rangle=\sqrt{J(J+1)-M(M\pm1)}|J,M\pm1\rangle$
with each $\hat{j}_{i}^{\pm}$ satisfying an analogous relation within
the $i$th atom. The emission rate is then given by $W=\sum_{M}\rho_{M}W_{J}(M)$,
where $\rho_{M}$ is the probability for the state being at the $M$th
level, and the associated collective decay rate is $W_{J}(M)=\gamma\langle\hat{D}^{+}\hat{D}^{-}\rangle_{JM}$
with $\gamma$ denoting the bare rate in free space.

For spin-$\nicefrac{1}{2}$ particles, the ladder operator $\hat{j}_{i}^{\pm}$
happens to be, up to a constant factor, equivalent to the dipole operator,
$\sigma_{i}^{-}=|g\rangle_{i}\langle e|$ and $\sigma_{i}^{+}=|e\rangle_{i}\langle g|$.
This connection makes it straightforward to obtain $\langle\hat{D}^{+}\hat{D}^{-}\rangle_{JM}=(J+M)(J-M+1)$
\cite{PhysRep82_Gross}. For spin-$j>\nicefrac{1}{2}$ atoms, the
spin and dipole operators are no longer parallel. To obtain an explicit
relation in this case, we take the mean-field assumption and get
\begin{eqnarray}
\langle\hat{D}^{+}\hat{D}^{-}\rangle_{JM} & = & \sum_{i}\langle\sigma_{i}^{+}\sigma_{i}^{-}\rangle+\sum_{i\neq j}\langle\sigma_{i}^{+}\sigma_{j}^{-}\rangle\nonumber \\
 & = & N\langle\sigma_{1}^{+}\sigma_{1}^{-}\rangle_{JM}+N(N-1)\langle\sigma_{1}^{+}\sigma_{2}^{-}\rangle_{JM},
\end{eqnarray}
where $\langle\sigma_{1}^{+}\sigma_{1}^{-}\rangle_{JM}$ and $\langle\sigma_{1}^{+}\sigma_{2}^{-}\rangle_{JM}$
can be further expressed in terms of the Clebsch-Gordan coefficients:
\begin{equation}
\langle\sigma_{1}^{+}\sigma_{1}^{-}\rangle=1-\langle j,(N-1)j;-j,M+j|J,M\rangle^{2}\label{eq:d1d1}
\end{equation}
and \begin{widetext}
\begin{eqnarray}
\langle\sigma_{1}^{+}\sigma_{2}^{-}\rangle & = & \sum_{m_{1},m_{2}}\Big[\langle j,(N-1)j;m_{1},M-m_{1}|J,M\rangle\langle j,(N-2)j;m_{2},M-m_{1}-m_{2}|J,M-m_{1}\rangle\times\nonumber \\
 &  & \langle j,(N-1)j;m_{1}-1,M-m_{1}+1|J,M\rangle\langle j,(N-2)j;m_{2}+1,M-m_{1}-m_{2}|J,M-m_{1}+1\rangle\Big].\label{eq:d1d2}
\end{eqnarray}
\end{widetext} The equation of motion now reads
\begin{eqnarray}
\dot{\rho}_{M=J} & = & -W_{J}(J)\rho_{J},\nonumber \\
\dot{\rho}_{M<J} & = & -W_{J}(M)\rho_{M}+W_{J}(M+1)\rho_{M+1}.\label{eq:dicke_eom}
\end{eqnarray}

\begin{figure}
\begin{centering}
\includegraphics[width=6.5cm]{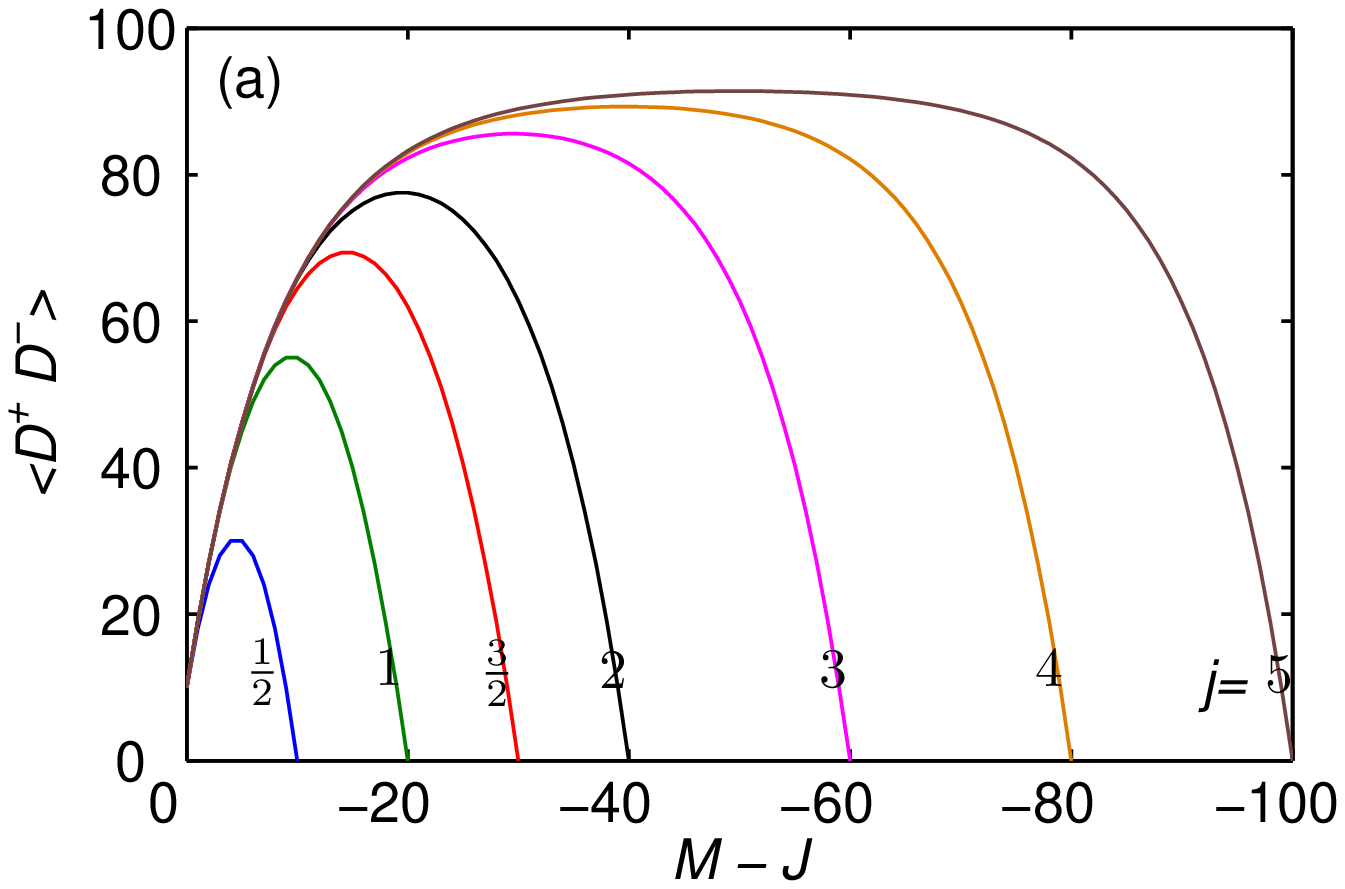}
\par\end{centering}

\begin{centering}
\includegraphics[width=6.5cm]{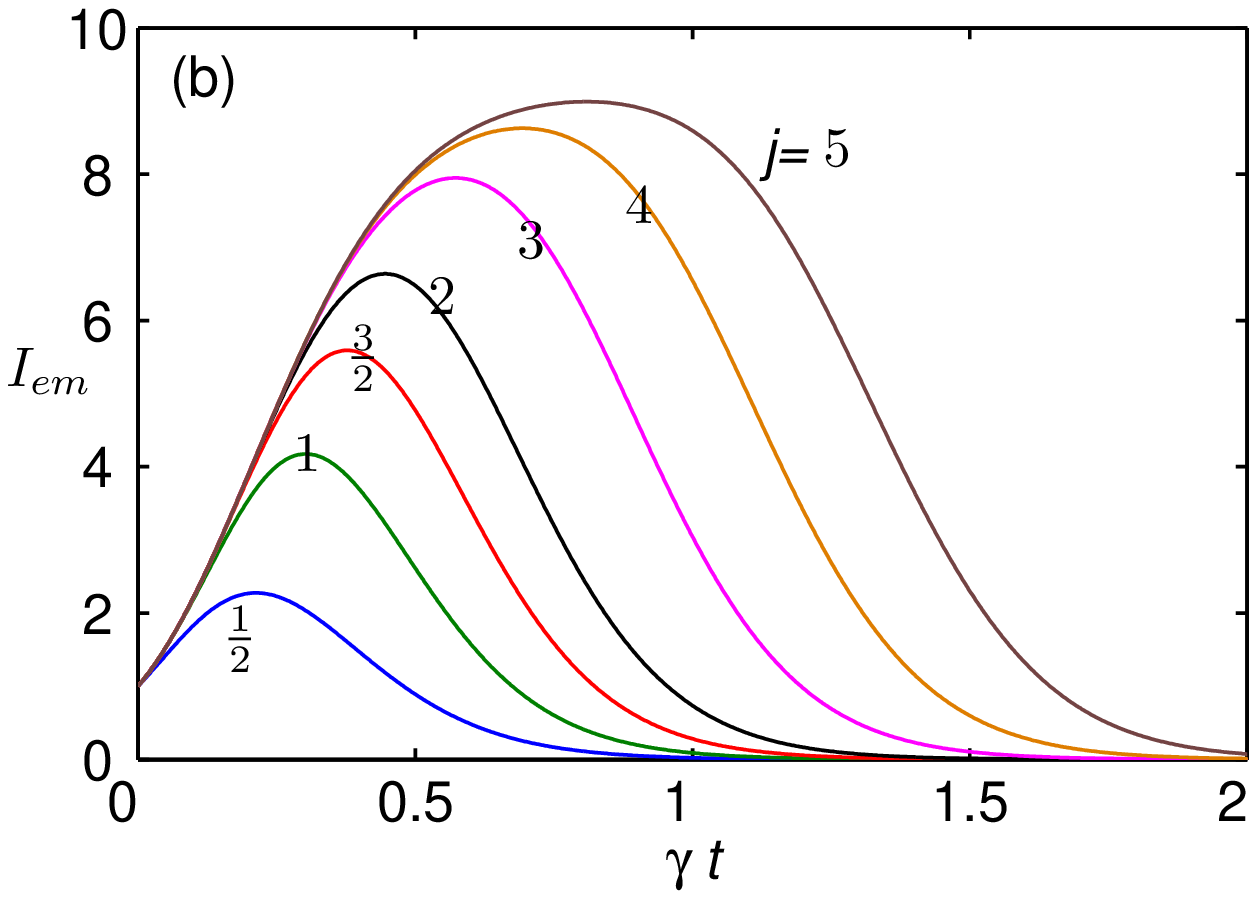}
\par\end{centering}

\begin{centering}
\includegraphics[width=6.5cm]{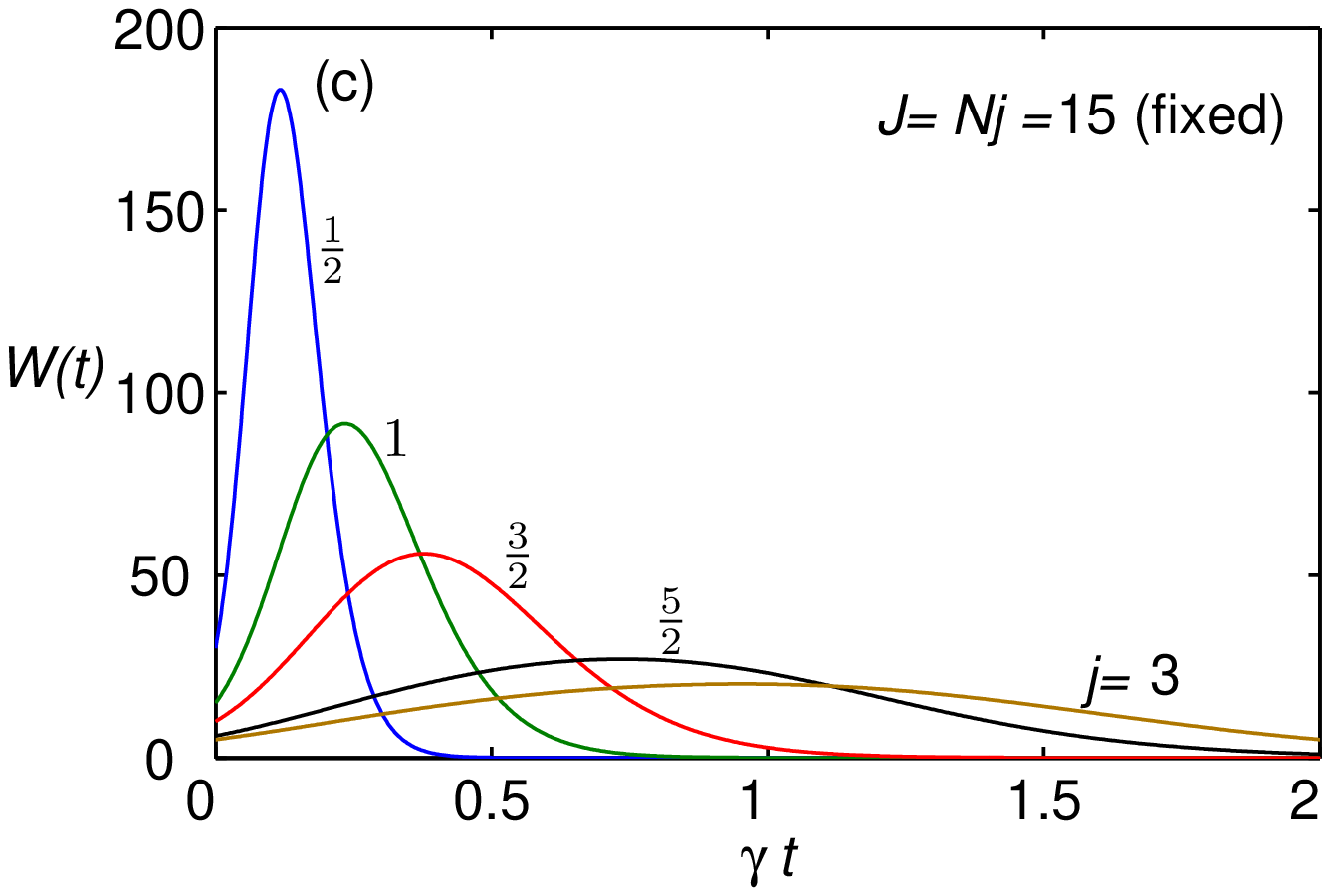}
\par\end{centering}

\caption{\label{fig:dickeresult}(color online). (a) $\langle\hat{D}^{+}\hat{D}^{-}\rangle_{JM}$
as a function of $M-J$ and (b) the emission intensity per particle
$I_{\text{em}}$ of different spin $j$'s for $N=10$ atoms; (c) Overall
emission curves for different spin $j$'s when $J=Nj=15$ is fixed.
Note that $I_{\text{em}}$ and $W$ are in units of $\gamma\hbar\omega_{0}$
with the energy level spacing $\hbar\omega_{0}$.}
\end{figure}

The emission curves of different $j$ are shown in Fig. \ref{fig:dickeresult}(b),
for which we calculate the intensity per particle $I_{\text{em}}=W(t)/N$
with $N=10$ by evolving Eq. (\ref{eq:dicke_eom}). One can observe
that every curve shows a different degree of superradiance behavior,
i.e., the intensity grows and maximizes in a short period of time.
As $j$ increases, the peak intensity becomes higher. This implies
that the radiation enhancement not only comes from many-body effects
but also from \emph{multiple levels}. This can be also seen in Fig.
\ref{fig:dickeresult}(a), where the enhancement factor $\langle\hat{D}^{+}\hat{D}^{-}\rangle_{JM}$
as a function of $M-J$ is plotted, consistently explaining the higher
emission rate for larger $j$. One feature worth noting is that all
$j$-curves in Fig. \ref{fig:dickeresult}(b) share the same {}``growing''
behavior. This is also suggested by the $\langle\hat{D}^{+}\hat{D}^{-}\rangle_{JM}$
curves in Fig. \ref{fig:dickeresult}(a): For we start the radiation
process from the fully excited state (the highest collective level),
the dynamics is dominantly determined by the population flows associated
with a few higher levels. The rates are proportional to the enhancement
factor $\langle\hat{D}^{+}\hat{D}^{-}\rangle_{JM}$. For various $j$,
we find in Fig. \ref{fig:dickeresult}(a) that these curves coincide
on the left hand side, corresponding to $|M-J|$ small (highest levels).
Another noticeable feature is that the $\langle\hat{D}^{+}\hat{D}^{-}\rangle_{JM}$
curves develop a plateau as $j$ increases and the highest value is
found to be bounded in the large-$j$ limit. It can then be expected
that the peak intensity value should also have a bounded value for
very large $j$. We will find this observation still true when we
use a more sophisticated approach. More details will be discussed
in Sec. \ref{sec:formalism}.

Finally, we compare the overall radiation by different spin-$j$ atoms,
keeping the total spin $J=Nj$ fixed. Among these cases, they share
the same $J$-Bloch sphere and therefore might be expected to have
similar behaviors. However this is not true, as we can see from Fig.
\ref{fig:dickeresult}(c). Smaller $j$ cases have faster and more
intense burst of emission while larger $j$ have smoother emission
rates and longer tails. This is because $\langle\hat{D}^{+}\hat{D}^{-}\rangle_{JM}$,
as determined by Eqs. (\ref{eq:d1d1}) and (\ref{eq:d1d2}), is also
dependent on $N$, not merely dependent on the total spin $J$.

\section{Effective two-body formalism\label{sec:formalism}}

We would like to emphasize that dipole-dipole interaction plays a
crucial role and is responsible for both real and virtual photon exchange.
As a consequence, the system builds up inter-particle coherence while
decaying. As dipole-dipole interaction is contained in Dicke's picture
in a sense that the spin flip-flops count, this picture treats the
whole system as a point-like object so that the inter-dipole coupling
is considered uniform. In an actual laboratory setup, Dicke's picture
is usually an oversimplified view because a real sample always occupies
a finite size and sees a finite wavelength of the radiative field.
The spatial arrangement of particles usually breaks permutation symmetry.
(Or more precisely, each particle sees different dipole-dipole couplings
to all others.) The nonuniform coupling leads to dephasing of the
Dicke states, resulting in population leakage out of the fully symmetric
manifold. Furthermore, dipole-dipole interaction also causes other
effects, e.g. frequency chirping, for which each Dicke state $|JM\rangle$
can be dipole-dipole shifted differently so that the emission frequency
becomes variable over time \cite{PhysRep82_Gross}. Superradiant behavior
becomes more complex (and less pronounced) when these effects are
not excluded. In order to better describe practical situations, we
need to go over the microscopic details of atom-field interactions.
The calculation, however, becomes intractable when the number of particles
increases typically for $N\gtrsim10$. In \cite{PRA99_Fleischhauer,Yelin05,PRA07_Wang},
we circumvent this difficulty by explicitly writing down the master
equation of motion for only two probe atoms, taking average over the
background atoms and tracing out the field variables. We also assume
that the field instantaneously interacts with the whole ensemble,
ignoring the retarding effects due to the finite size. (This can be
justified because the characteristic time of propagation $d/c$ ($\sim10^{-12}$
sec for a sample of size $d\sim1$ mm) is usually much smaller than
any other decay timescales.) We summarize here the main results and
leave the details of the derivation to the Appendix. The relevant
two-body master equation is given by 
\begin{eqnarray}
\dot{\rho} & = & -\sum_{i,j=1,2}\frac{\Gamma_{ij}}{2}\big(\big[\rho\sigma_{i}^{-},\sigma_{j}^{\dagger}\big]+\big[\sigma_{i}^{-},\sigma_{j}^{\dagger}\rho\big]\big)\nonumber \\
 & - & \sum_{i,j=1,2}\frac{\Gamma_{ij}+\gamma\delta_{ij}}{2}\big(\big[\rho\sigma_{j}^{\dagger},\sigma_{i}^{-}\big]+\big[\sigma_{j}^{\dagger},\sigma_{i}^{-}\rho\big]\big),\label{eq:mastereq}
\end{eqnarray}
where $\rho$ is the two-body density matrix with dimension $(2j+1)^{2}\times(2j+1)^{2}$,
$\gamma=\frac{\wp^{2}\omega_{0}^{3}}{3\pi\hbar\epsilon_{0}c^{3}}$
is the free-space spontaneous decay rate, $\Gamma\equiv\Gamma_{ii}$
is the single-particle induced pump/decay rate, and $\bar{\Gamma}\equiv\Gamma_{ij}$
($i\neq j$) denotes the two-particle damping rate responsible for
the atom-atom correlation. The mean-field approximation with the second
order correction in fields gives the self-consistent form for the
induced rate:
\begin{eqnarray}
\Gamma & = & \gamma(e^{2\zeta}-1)\frac{A(t)}{V(t)}+2\mathcal{C}^{2}\varrho^{4}\frac{\gamma^{2}I(\zeta,\varrho)}{\Gamma+\gamma/2}Y(t)\label{eq:Gamma1}\\
\bar{\Gamma} & = & \frac{\gamma^{2}I(\zeta,\varrho)}{\Gamma+\gamma/2}\left[3\mathcal{C}\varrho A(t)+2\mathcal{C}^{2}\varrho^{4}Y(t)\right],\label{eq:Gamma2}
\end{eqnarray}
with
\begin{eqnarray}
A(t) & = & \sum_{m=-j+1}^{j}\rho_{mm}^{(1)}\label{eq:At}\\
V(t) & = & \rho_{jj}^{(1)}-\rho_{-j,-j}^{(1)}\label{eq:Vt}\\
Y(t) & = & \sum_{m,m^{\prime}=-j}^{j-1}\rho_{m+1,m;m^{\prime},m^{\prime}+1},\label{eq:Yt}
\end{eqnarray}
where $\rho^{(1)}\equiv\frac{1}{2}\sum_{i=1,2}\text{tr}_{i}[\rho]$
denotes the reduced single-particle density matrix and $\rho_{ab;cd}\equiv\frac{1}{2}[\langle a,c||\rho|b,d\rangle+\langle c,a|\rho|d,b\rangle]$.
The factor $\frac{1}{2}$ comes from averaging the interchanging of
two particles. Note that interchange symmetry requires $\rho_{ab;cd}=\rho_{cd,ab}$
and $\rho_{ab;cd}^{\ast}=\rho_{ba;dc}$. The cooperativity parameter
is defined as $\mathcal{C}\equiv2\pi c^{3}\mathcal{N}/\omega^{3}$;
$\varrho\equiv\omega d/(2c)$ characterizes the system size $d$ in
terms of the radiation wavelength. These results are based on assumptions
that there is no external field and hence the generated field has
to be on-resonant with the transition frequency. Function $I(\zeta,\varrho)\equiv[\big((\zeta-1)e^{\zeta}+\cos\varrho\big)^{2}+\big(\varrho e^{\zeta}-\sin\varrho\big)^{2}]/(\zeta^{2}+\varrho^{2})^{2}\approx\frac{e^{2\zeta}}{\zeta^{2}+\varrho^{2}}$
for large $\zeta$ and $\varrho$. If no thermal broadening is assumed,
we have $\zeta\equiv\frac{1}{2}\mathcal{C}\varrho\frac{\gamma}{\Gamma+\gamma/2}V(t)$.
When the Doppler broadening needs to be considered, the fields allow
to be detuned and these quantities must be averaged. More details
will be discussed in Sec. \ref{sec:Doppler}.

\section{Results\label{sec:result}}

\subsection{Emission and decay rates}

\begin{figure}
\begin{centering}
\includegraphics[width=7cm]{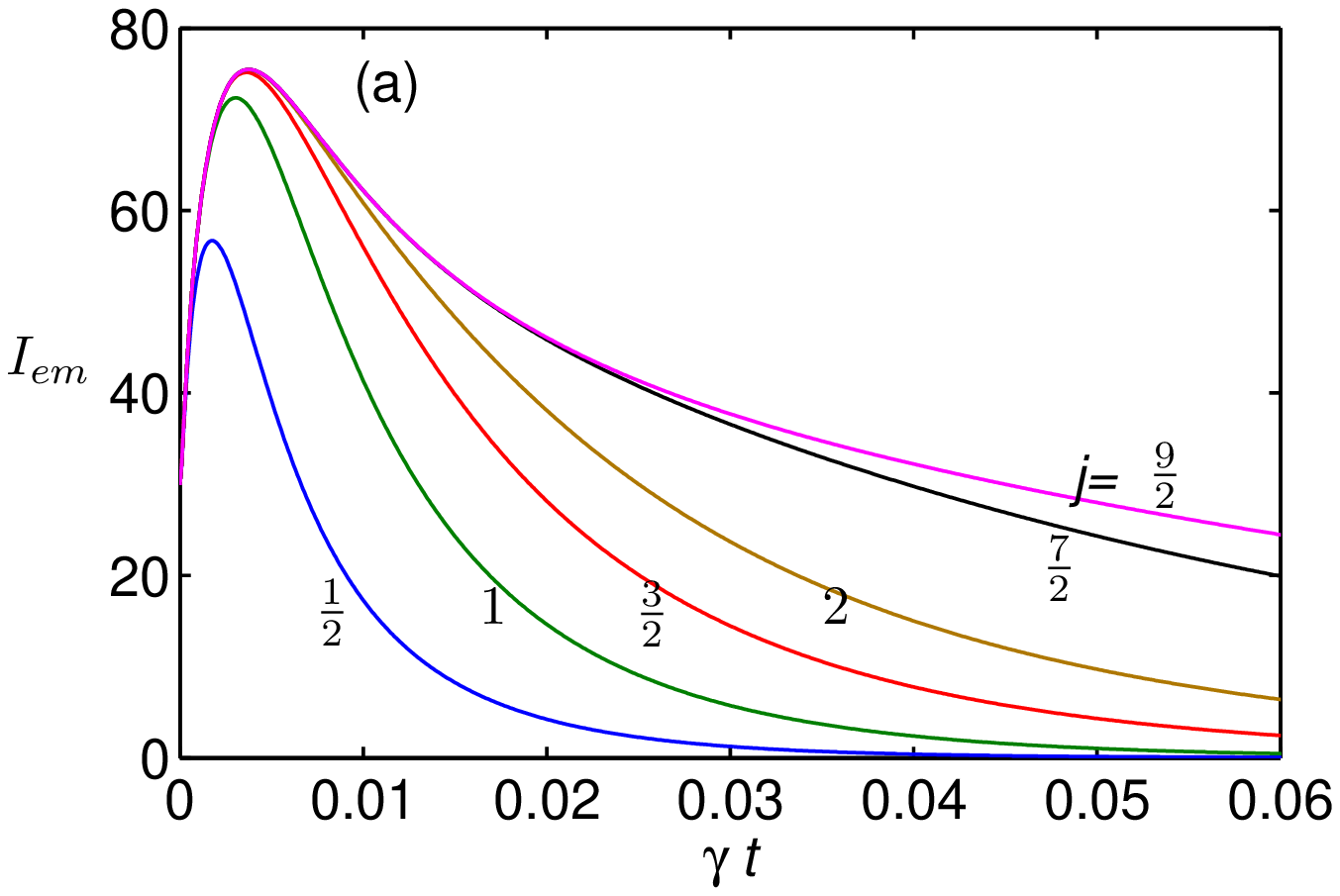}
\par\end{centering}

\begin{centering}
\includegraphics[width=7cm]{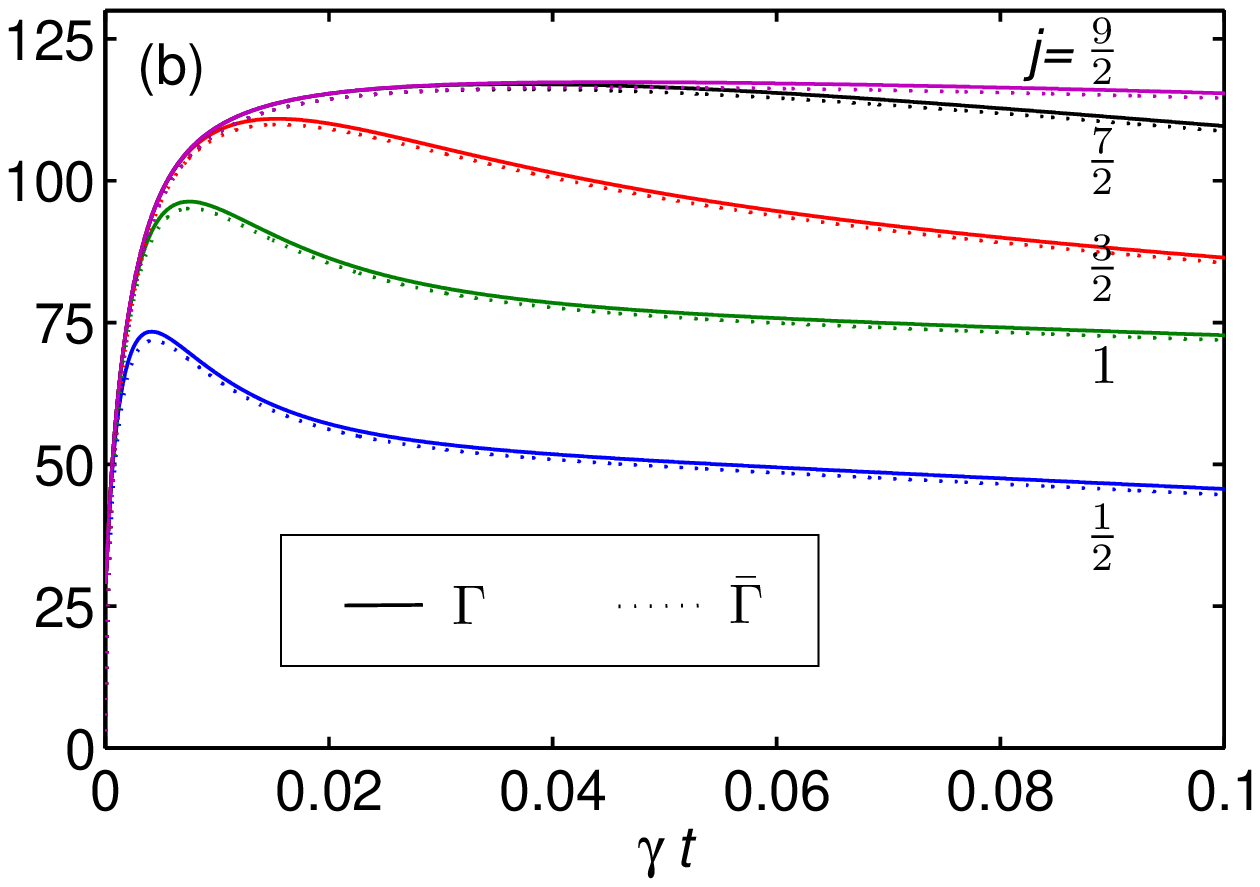}
\par\end{centering}

\caption{\label{fig:spinj}(color online). The temporal profiles of (a) the
emission intensity per particle (in units of $\gamma\hbar\omega_{0}$)
and (b) the induced single-atom pump/decay rate $\Gamma$ and two-atom
correlation damping rate $\bar{\Gamma}$ (in units of $\gamma$).
In all cases we use $\mathcal{C}=10$ and $\varrho=10$. }
\end{figure}

By Eq. (\ref{eq:mastereq}) we are able to solve for the temporal
emission rate curve. Fig. \ref{fig:spinj}(a) shows the radiation
intensity per particle $I_{\text{em}}$ with $\mathcal{C}=10$ and
$\varrho=10$ for different spin-$j$ species, where $I_{\text{em}}\equiv\hbar\omega_{0}\sum_{m=-j}^{j}(j+m)\frac{d}{dt}\rho_{mm}^{(1)}(t)$.
Here we take $\rho_{jj,jj}=1$ and set $0$ for all other density
matrix elements as the initial state. It can be seen that for each
$j$ the radiation intensity reaches a peak, giving a strong evidence
of superradiance. Those curves follow roughly the same intensity profile
in the beginning. The maximal value of intensity first grows as $j$
increases from $\nicefrac{1}{2}$, and then stops growing when $j\gtrsim2$.
The time of reaching the peak intensity also converges to a fixed
constant $t_{\text{max}}$ in the large $j$ limit. This also has
been observed in the Dicke superradiance picture. Here we give an
intuitive explanation as follows: When the system starts to relax
from the state with all atoms initially excited to the highest level,
only a few highest levels are involved in determining the radiative
behavior during the early stage. Even for a very large spin particle,
who has a huge multi-level structure, those levels lower than the
first few have not been populated yet and hence do not have contributions.
The time evolution of the decay rates is also plotted in Fig. \ref{fig:spinj}(b).
Note that at $t=0$ the diagonal decay rate $\Gamma(0)$ determines
the initial emission intensity, followed by a sharp growth and hence
resulting in intensity peaks. In the mean time, the off-diagonal $\bar{\Gamma}$
emerges and mixes single-body states. Different from the Dicke picture,
where we choose the eigenbasis to be the symmetric states constructed
by a giant spin object $J=Nj$, here we use products of single particle
states as the eigenbasis, allowing the degrees of freedom of population
being transferred to asymmetric levels. Note that the dipole-dipole
interaction is built-in in our formalism and is responsible for these
effects. Consequently, the superradiance enhancement with $j$, when
characterized by the growth of peak intensity, in more realistic cases
cannot be as large as predicted by the Dicke model. On the other hand,
since the asymmetric levels have lower or vanishing decay rates, the
occupation of these levels modifies the tails of the emission curves.
In some circumstances, the energy is trapped. Such effects cannot
be described by the Dicke model.

A few remarks are placed here regarding the connection of cooperativity
and superradiant curves. As we increase $\mathcal{C}$ ($\varrho$)
while fixing $\varrho$ ($\mathcal{C}$), the emission peak intensity
per particle increases proportionally while the time scale of the
initial intensity burst is inversely proportional to $\mathcal{C}$
($\varrho$). This is due to the {}``many-body enhancement'', as
we have discussed using Dicke's picture. Such features are commonly
observed even in the original two-level systems. Furthermore, the
emission curves are found to be similar when $\mathcal{C}\varrho$
is kept the same (not shown). This can also be seen analytically (see
Appendix). This suggests that $\mathcal{C}\varrho\sim\mathcal{N}\lambda^{2}d$
be the relevant factor that determines the primary superradiant behavior
while the detailed emission curves, however, still slightly depend
on $\mathcal{C}$ and $\varrho$ separately \cite{PRL08_Akkermans,PRA07_Wang}.

\subsection{Significance of atom-atom coherence}

\begin{figure}
\begin{centering}
\includegraphics[width=7cm]{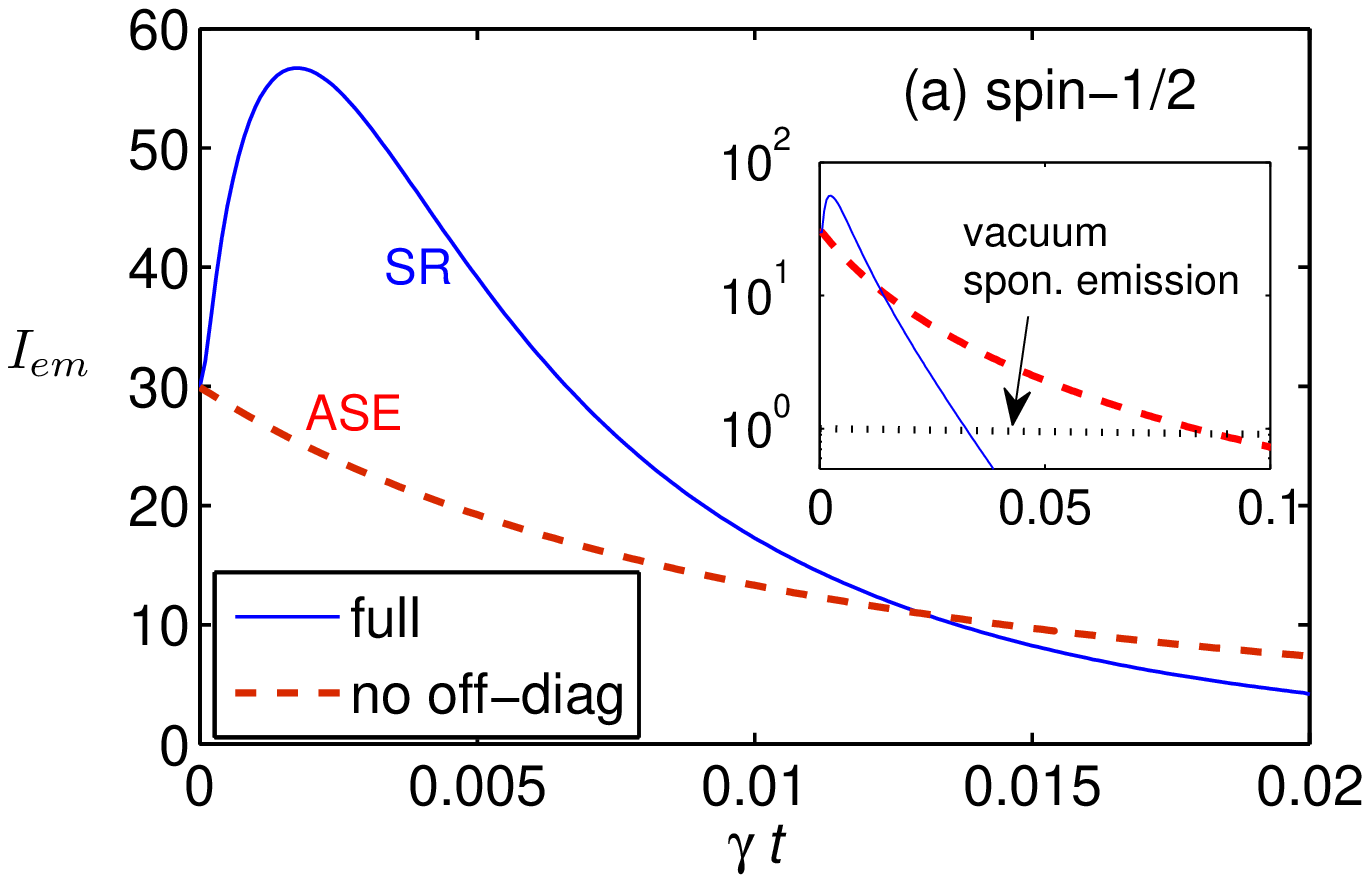}
\par\end{centering}

\begin{centering}
\includegraphics[width=7cm]{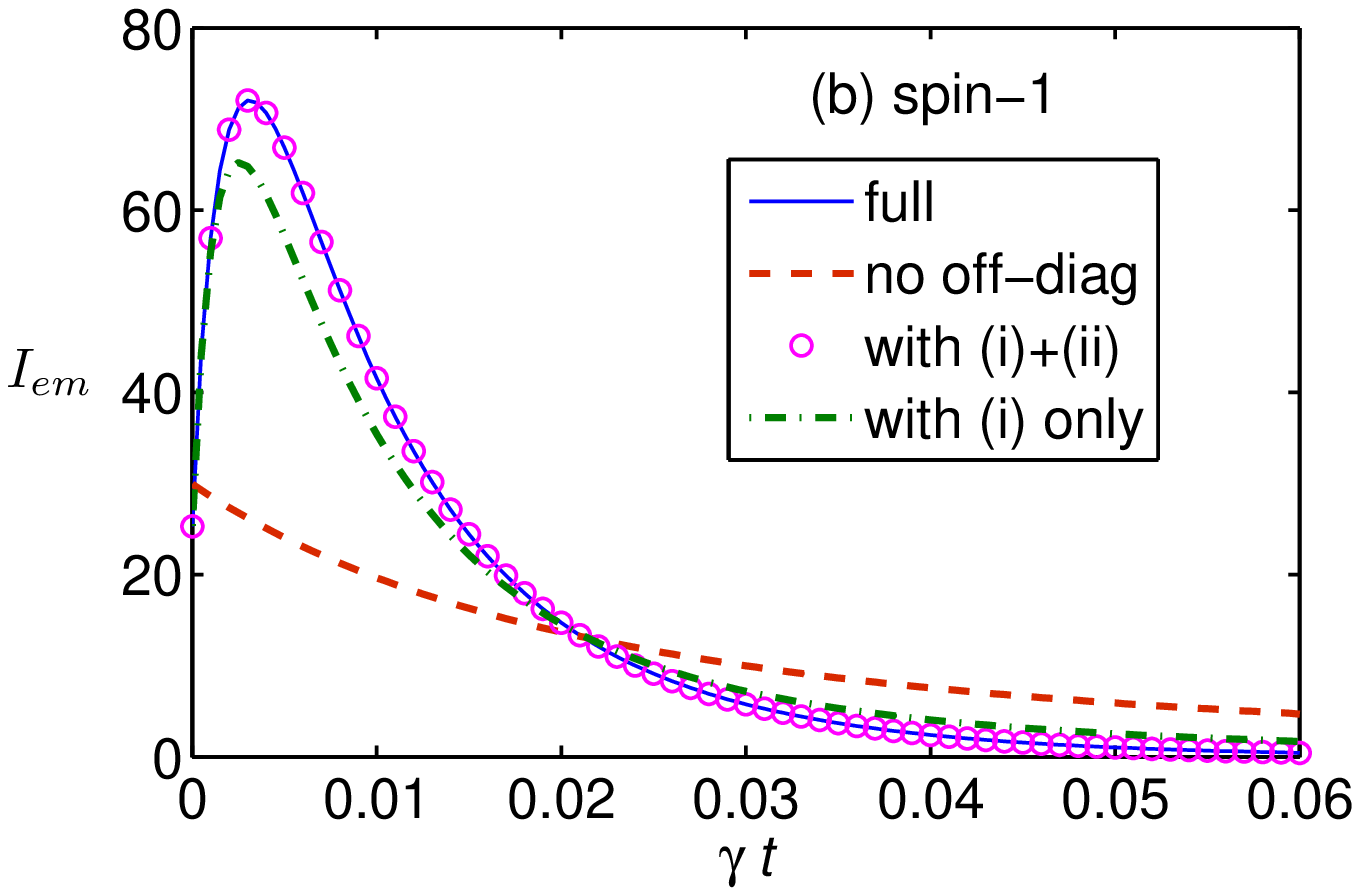}
\par\end{centering}

\begin{centering}
\includegraphics[width=7cm]{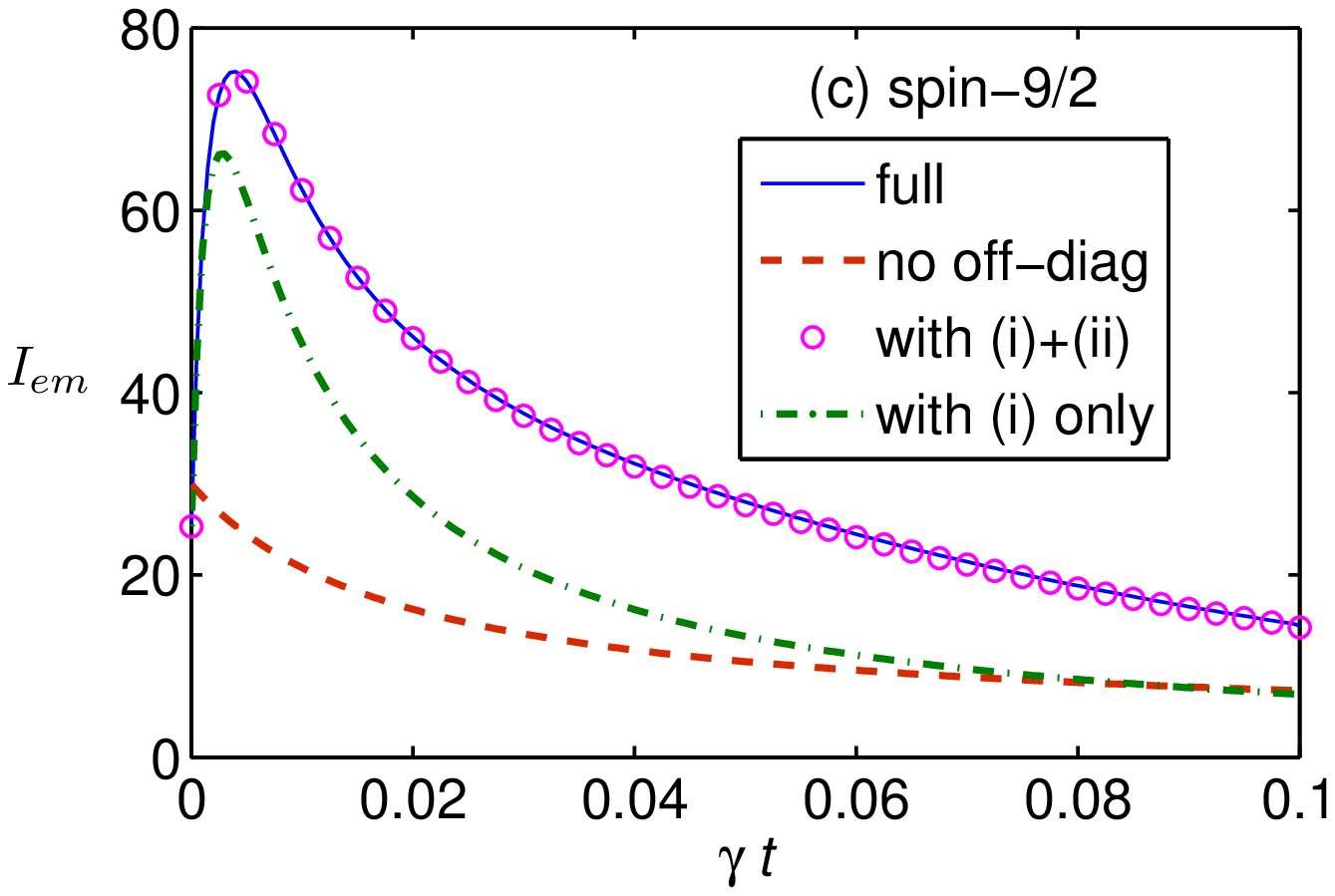}
\par\end{centering}

\caption{\label{fig:wo_coh}(color online). Evolution curves for emission intensity
when the off-diagonal terms are fully considered (full), partially
removed (with both (i) and (ii), or only with (i); see text), or entirely
removed (no off-diag) for (a) spin-$\nicefrac{1}{2}$, (b) spin-$1$,
and (c) spin-$\nicefrac{9}{2}$ particles. Other parameters are the
same as Fig. \ref{fig:spinj}. Inset of (a): We plot three curves
of superradiance (SR, blue solid), amplified spontaneous emission
(ASE, red dashed), and single-particle free-space spontaneous emission
(black dotted) for comparison.}
\end{figure}

Cooperation of many-body states is crucial to superradiance. The beauty
of this formalism is that we retain the accessibility to atom-atom
correlations under the framework of the mean field approximation.
Here we investigate the role of many-body correlations, which, in
our case, are contained in the off-diagonal terms of the two-body
density matrix. To see this, this method allows us to manipulate these
off-diagonal terms and evolve Eq. (\ref{eq:mastereq}). The results
will then be compared. Note that these off-diagonal terms have the
form $\rho_{a,a+m;b,b-m}$. In the spin-$\nicefrac{1}{2}$ case, the
only possibility is $\rho_{eg,ge}$. Fig. \ref{fig:wo_coh}(a) shows
the emission curve for $\mathcal{C}=10$, $\varrho=10$, and $\rho_{eg,ge}=0$
at all time. This curve is now found to be monotonically descending,
signaling mere amplified spontaneous emission (ASE) instead of superradiance
(SR) because of apparent lack of an intensity peak. Such monotonicity
is shared by larger $j$ cases (Figs. \ref{fig:wo_coh}(b) and (c))
when all off-diagonal terms are set to zero. The reason is clear:
Without atom-atom correlation, the density matrix is reduced to a
single-particle description and therefore no cooperative effects are
observed. On the other hand, for larger $j$ atoms the off-diagonal
terms does not only concern atom-atom correlations from the same-level
transitions but also involve that from different-level density matrix
elements. In the following we try to distinguish the importance from
three kinds of coherence terms: (i) the same-level coherence $\rho_{a,a+1;a+1,a}$,
(ii) the cross-coherence $\rho_{a,a+1;b+1,b}$ for $a\neq b$, and
(iii) the higher-order coherence $\rho_{a,a+m;b,b-m}$ for $m\ge2$.
For example, in Figs. \ref{fig:wo_coh}(b) for spin-$1$ and (c) for
spin-$\nicefrac{9}{2}$ atoms, we plot the emission curves corresponding
to all off-diagonal terms being dropped out, and (i), (ii), and then
(iii) being added back to the system. It can be found that by inclusion
of (i) and (ii) the system has already behaved like the actual dynamics,
indicating that the higher-order coherence is negligible in determining
the evolution of the system. However, if only (i) is included, although
the intensity enhancement can still be observed, the details of emission
profile have discrepancy to the actual behaviors. The distinction
becomes even more obvious when $j$ gets large, as we see in Fig.
\ref{fig:wo_coh}(c). The fact that the cross-coherence terms must
be taken into consideration implies the interferences due to {}``cross-level''
transitions, i.e., the transitions of the same energy difference,
but not from two definite levels, have some kind of {}``multi-level''
contributions to the superradiance, in analog to the many-body effects.

\section{Doppler broadening\label{sec:Doppler}}

When a hot gas is considered, the energy difference seen by moving
atoms varies. In this section we consider the loss of coherence due
to the Doppler effects. Suppose that the thermal gas is described
by a Gaussian distribution function: 
\begin{equation}
f_{D}(\delta)=\frac{1}{\sqrt{2\pi}\Delta_{D}}\exp\Big[-\frac{\delta^{2}}{2\Delta_{D}^{2}}\Big],\label{eq:Gaussian_distribution}
\end{equation}
where $\delta$ is the Doppler shift for an atom and $\Delta_{D}$
is the characteristic width of this distribution. In order to take
this average into account we need to go to the derivation summarized
in the Appendix. Note that the frequency difference between the field
and the Fourier component $\Delta$ in Eqs. (\ref{eq:Gamma1_integration},\ref{eq:Gamma2_integration})
is now modified to $\Delta-\delta$. Here we also assume that the
velocity distribution will not be affected by atoms' recoil when photons
are emitted. The idea is then to divide the whole system in the frequency
space into a series of small slivers according to the detunings they
see individually. For each sliver, it can be expected that the integration
as in Eqs. (\ref{eq:Gamma1_integration},\ref{eq:Gamma2_integration})
should lead to the same spatial dependence. The only modification
is that the source and retarded functions (\ref{eq:source_P1},\ref{eq:source_P2},\ref{eq:retarded_P})
and Green's functions (\ref{eq:retarded_D}) need to be determined
in an averaged manner because they contain the effects from all slivers.
For convenience we use a notation that $\bar{Q}=\int_{-\infty}^{\infty}Q(\delta)f_{D}(\delta)d\delta$
to represent an Doppler averaged quantity. Consequently, 
\begin{eqnarray}
\overline{\widetilde{P}^{ret}}(\Delta) & = & \frac{\mathcal{N}\wp^{2}}{\hbar^{2}}V(t)\overline{\frac{1}{\Gamma_{f}-i(\Delta-\delta)}}\nonumber \\
 & = & \frac{\mathcal{N}\wp^{2}}{\hbar^{2}}\frac{V(t)}{\Delta_{D}}\sqrt{\frac{\pi}{2}}U(iz_{0}),\\
\overline{\widetilde{P}^{s}}(\Delta) & = & \frac{\mathcal{N}\wp^{2}}{\hbar^{2}}\frac{2A(t)}{\Delta_{D}}\sqrt{\frac{\pi}{2}}\mbox{Re}[U(iz_{0})]\\
\overline{\widetilde{D}^{ret}}(\vec{x},\Delta) & = & -\frac{i\hbar\omega^{2}}{6\pi\epsilon_{0}c^{2}}\frac{e^{\overline{q_{0}^{\prime\prime}}x}}{x}e^{-i\overline{q_{0}^{\prime}}x},
\end{eqnarray}
where $U(z)\equiv\frac{2}{\sqrt{\pi}}\int_{z}^{\infty}e^{z^{2}-s^{2}}ds$
is the scaled complementary error function and $z_{0}=\frac{\Gamma+\gamma/2+i\Delta}{\sqrt{2}\Delta_{D}}$;
$\overline{q_{0}^{\prime\prime}}=\frac{C\gamma\varrho}{d}V(t)\frac{1}{\Delta_{D}}\sqrt{\frac{\pi}{2}}U(iz_{0})$
while $\overline{q_{0}^{\prime}}=q_{0}=\omega/c$ keeps the same.
Finally we have
\begin{eqnarray}
\Gamma(\Delta) & = & \gamma(e^{2\bar{\zeta}}-1)\frac{A(t)}{V(t)}\nonumber \\
 &  & +2\frac{\gamma^{2}}{\Delta_{D}}\mathcal{C}^{2}\varrho^{4}I(\bar{\zeta},\bar{\varrho})\mbox{Re}[U(iz_{0})]Y(t)\label{eq:Doppler_Gamma1}\\
\bar{\Gamma}(\Delta) & = & \frac{\gamma^{2}}{\Delta_{D}}I(\bar{\zeta},\bar{\varrho})\mbox{Re}[U(iz_{0})]\times\nonumber \\
 &  & \left[3\mathcal{C}\varrho A(t)+2\mathcal{C}^{2}\varrho^{4}Y(t)\right],\label{eq:Doppler_Gamma2}
\end{eqnarray}
where 
\begin{eqnarray}
\bar{\varrho}(\Delta) & = & \varrho+\frac{1}{2}\sqrt{\frac{\pi}{2}}\mathcal{C}\gamma\varrho\frac{V(t)}{\Delta_{D}}\text{Re}[U(iz_{0})]\label{eq:varrho_bar}\\
\bar{\zeta}(\Delta) & = & \frac{1}{2}\sqrt{\frac{\pi}{2}}\mathcal{C}\gamma\varrho\frac{V(t)}{\Delta_{D}}\text{Im}[U(iz_{0})].\label{eq:zeta_bar}
\end{eqnarray}
Each sliver, depending on its location at the thermal distribution
function, sees these quantities associated with a specific Fourier
component $\Delta$. The overall decay is then equivalent to averaging
$\Gamma$ and $\bar{\Gamma}$ according to the thermal distribution
function, i.e.,
\begin{equation}
\Gamma_{ij}^{D}=\int_{-\infty}^{\infty}d\Delta\frac{1}{\sqrt{2\pi}\Delta_{D}}e^{-\Delta^{2}/(2\Delta_{D}^{2})}\Gamma_{ij}(\Delta).\label{eq:Doppler_average}
\end{equation}

\begin{figure}
\begin{centering}
\includegraphics[width=7cm]{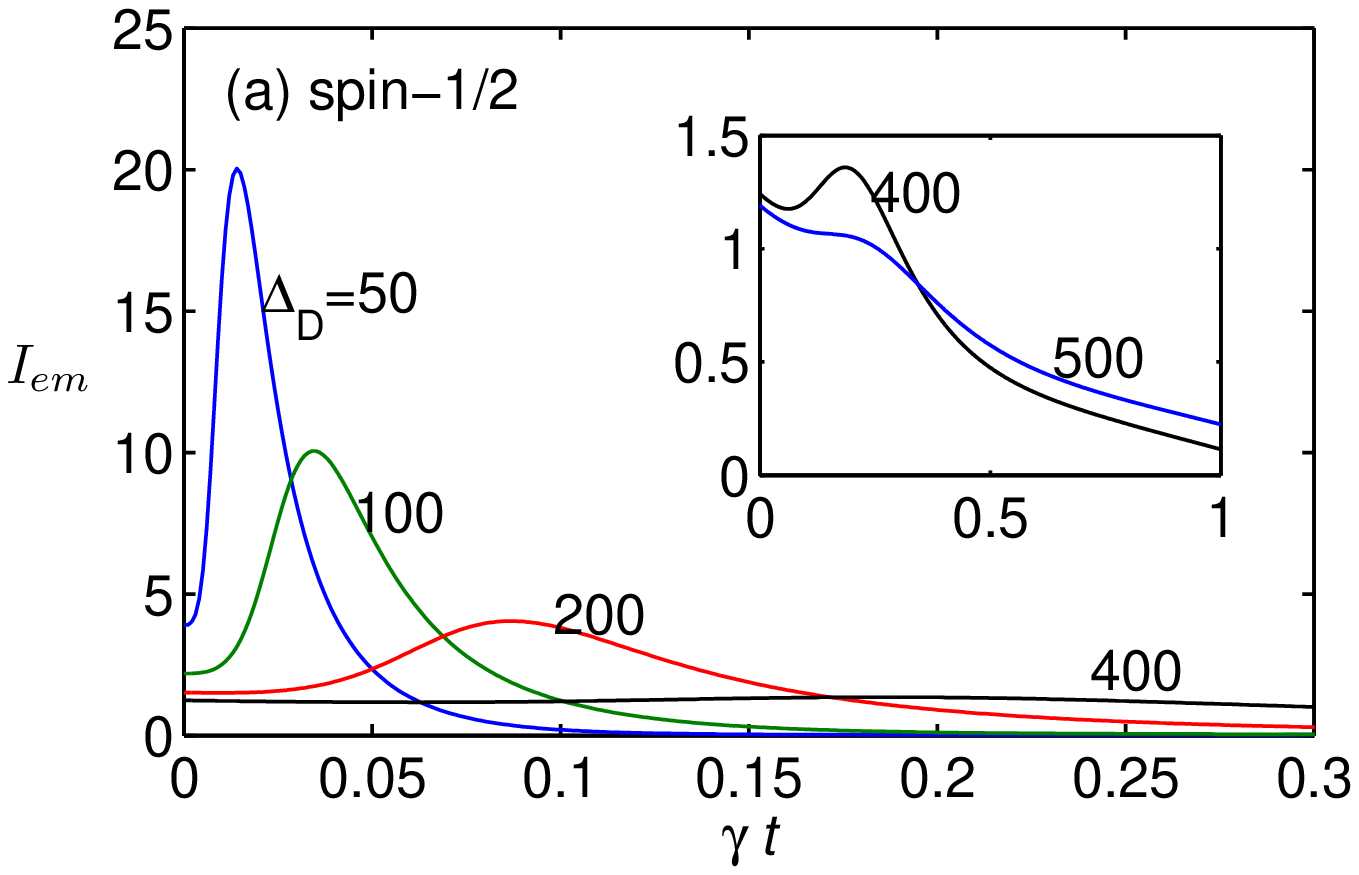}
\par\end{centering}

\begin{centering}
\includegraphics[width=7cm]{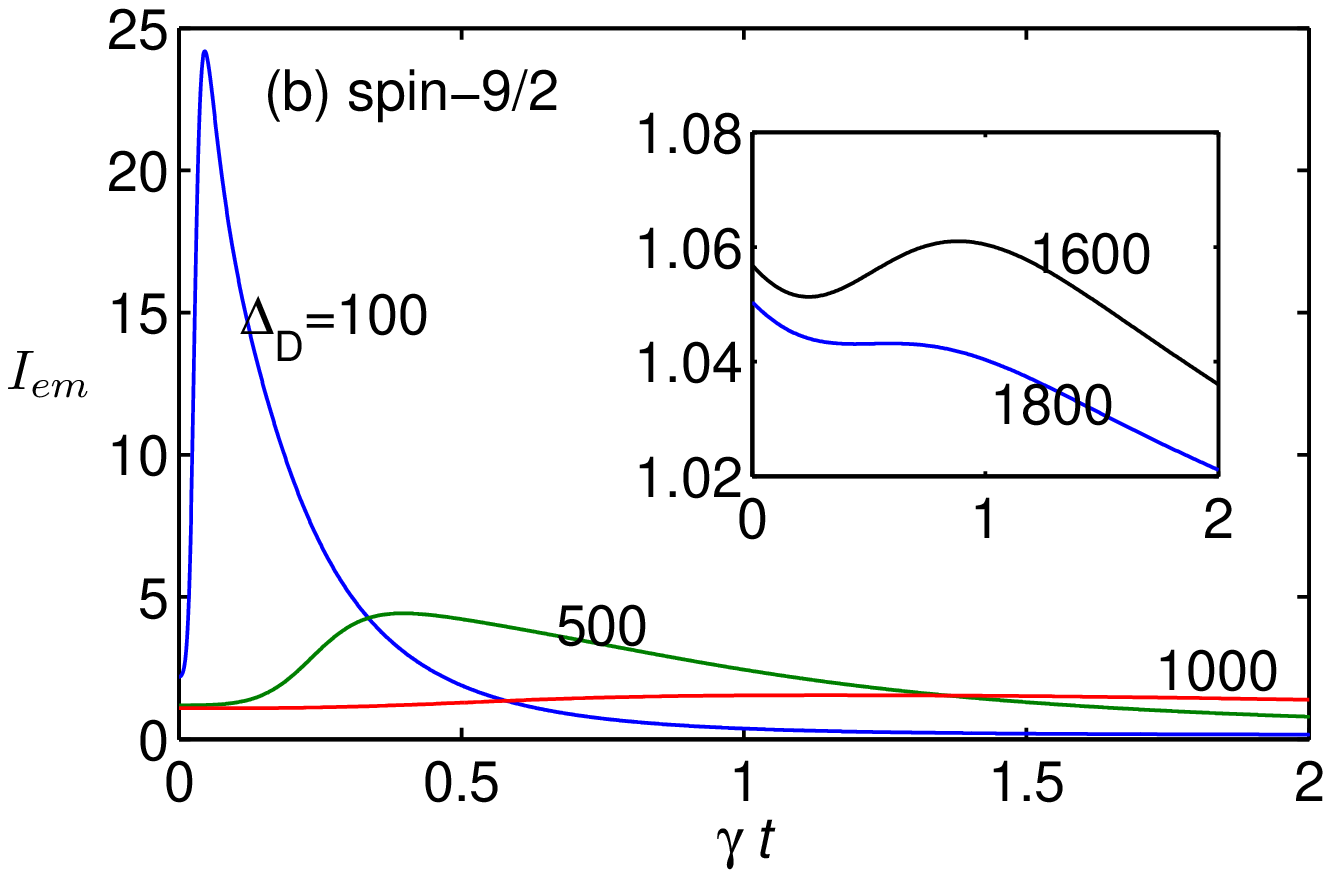}
\par\end{centering}

\caption{\label{fig:emission_doppler} (color online). Doppler-broadened emission
intensity curves for (a) spin-$\nicefrac{1}{2}$ and (b) spin-$\nicefrac{9}{2}$
particles. The inset figures show, for each case, two curves slightly
smaller and larger than the marginal Doppler width $\Delta_{m}$.
Here we use $\mathcal{C}=10$ and $\varrho=10$ for both cases, and
get (a) $\Delta_{m}/\gamma=433$ and (b) $\Delta_{m}/\gamma=1650$.}
\end{figure}

\begin{figure}
\begin{centering}
\includegraphics[width=7cm]{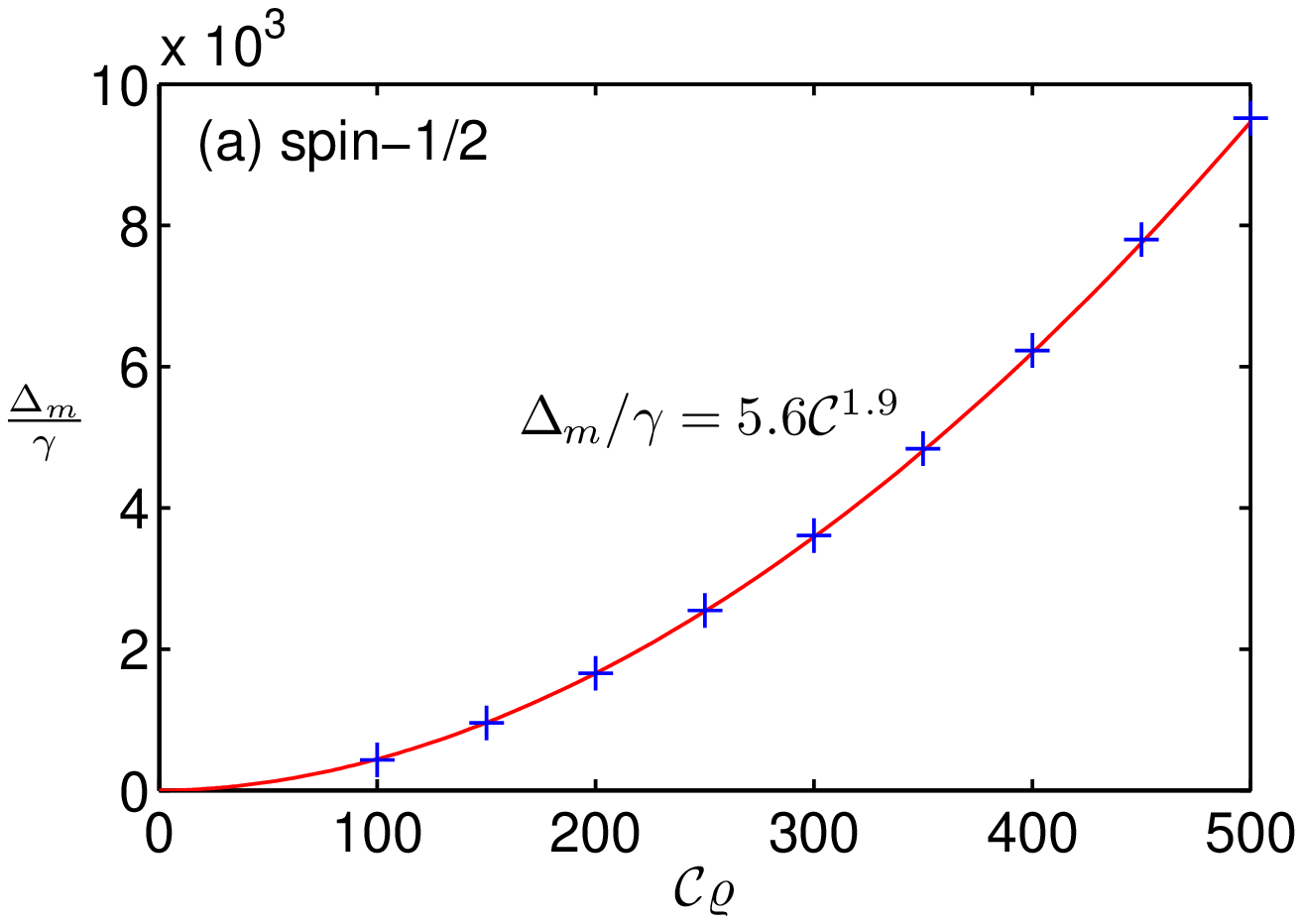}
\par\end{centering}

\begin{centering}
\includegraphics[width=7cm]{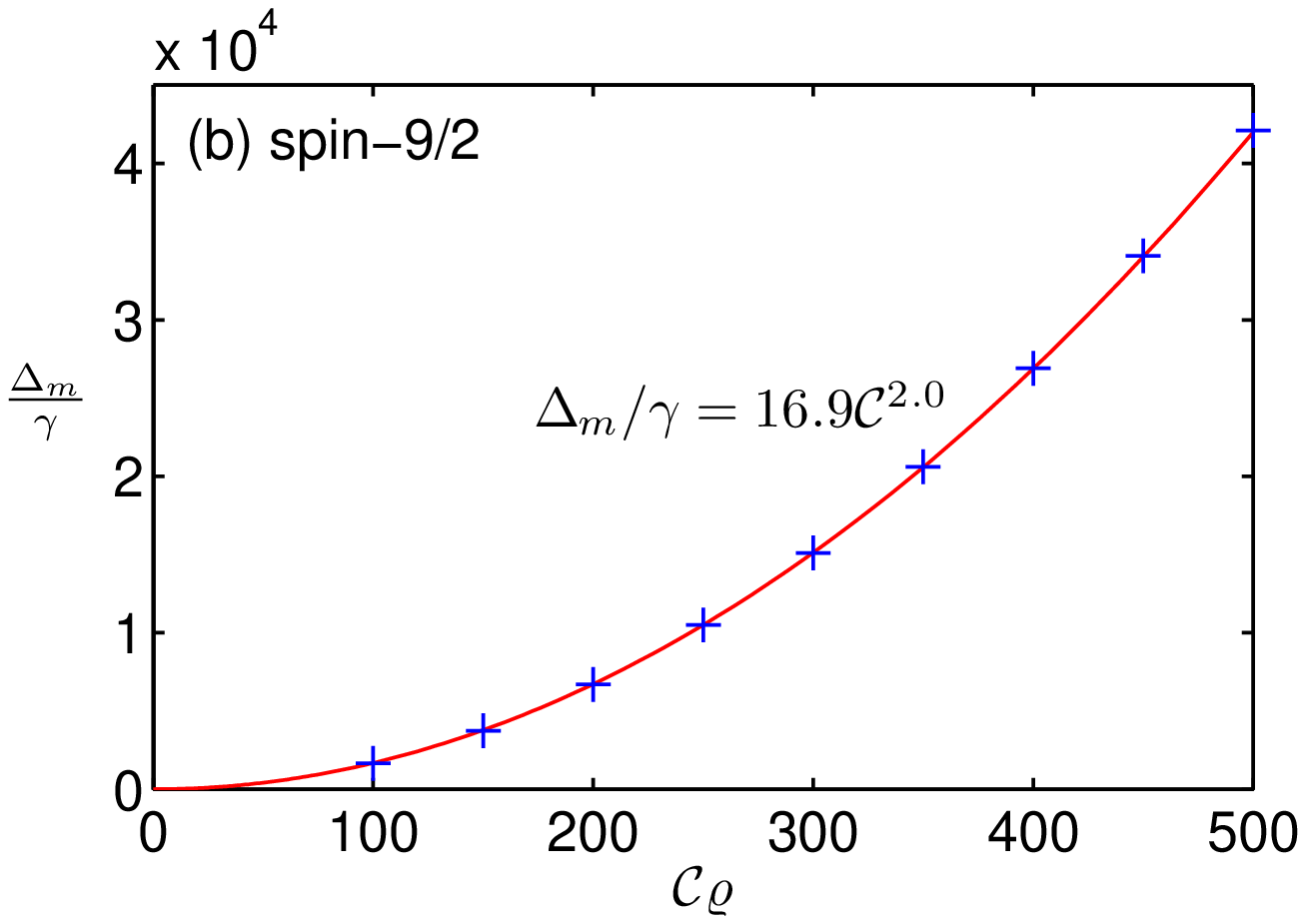}
\par\end{centering}

\caption{\label{fig:Delta_m}(color online). The marginal Doppler width $\Delta_{m}$
as a function of the cooperativity for (a) spin-$\nicefrac{1}{2}$
and (b) spin-$\nicefrac{9}{2}$ particles. (Here we take $\varrho=10$).
The cross dots (blue) represent our calculated data; the solid lines
(red) are best fitting power-law curves with the exponent around $2$.}
\end{figure}

Eq. (\ref{eq:Doppler_average}) is solved numerically. We then calculate
the corresponding emission curves and show them in Fig. \ref{fig:emission_doppler}.
With the Doppler broadening, it is clear that the superradiance behavior
is suppressed as the Doppler width $\Delta_{D}$ increases. This is
due to frequency mismatch and therefore part of the atoms loses track
of coherence and decays more independently. But the superradiance
peaks are still observable within a certain range of $\Delta_{D}$,
until it becomes too large and kills the peaks. This is understandable
because for a small-width distribution, there are still sufficient
atoms located within a frequency-matching regime. In order to characterize
the existence of superradiance when the match and mismatch parts compete,
we define a marginal Doppler width $\Delta_{m}$ beyond which the
peak value no longer surpasses the initial intensity. Obviously $\Delta_{m}$
depends on the cooperativity parameter $\mathcal{C}$, and in Figs.
\ref{fig:Delta_m}(a) and (b) we plot the relations for spin-$\nicefrac{1}{2}$
and spin-$\nicefrac{9}{2}$ particles, respectively. We find approximately
a quadratic dependence $\Delta_{m}\propto\mathcal{C}^{2}$. Note that
when the cooperativity parameter increases, not only the total number
of particle increases but the inter-particle spacing decreases accordingly.
This enhances the dipole-dipole interaction (of the order of magnitude
$\wp^{2}/(2\pi\epsilon_{0}r^{3})\propto\mathcal{C}$) which is responsible
for the superradiance for an additional factor $\mathcal{C}$. As
a result we have the tolerance $\Delta_{m}$ with a quadratic dependence
on $\mathcal{C}$ rather than a linear dependence.

\section{Molecular vibrational states \label{sec:vibration}}

One direct example for multi-level structure is vibrational modes
of polar molecules where the deeply bound potential can be well approximated
by a harmonic one. The number of low lying eigenstates that are quasi-equally
spaced energy levels can usually be up to a few tens. These particles
are thus analogous to large {}``spin'' particles. Take a typical
example of heteronuclear diatomic alkali molecules \cite{Deiglmayr11}:
the state $X^{1}\Sigma^{+}$ for LiCs has an averaged energy spacing
$\omega_{0}\approx2\pi\times5$ THz. For a sample of LiCs molecules
with a density $\mathcal{N}\approx4\times10^{9}$ $\mbox{cm}^{-3}$,
such energy spacing corresponds to cooperativity $\mathcal{C}\sim20\gg1$.
The transitional dipole moment between two adjacent vibrational states
is about $5$ Debye, and therefore the single-particle spontaneous
emission rate $\gamma\sim\mbox{sec}^{-1}$. We then expect within
this parameter regime that superradiance intensity peak can be observed
in a timescale of mini-sec while $\Gamma_{\text{max}}\gtrsim1000\gamma$.

The cascade relaxation of excited population from higher to lower
levels is a reminiscence of motional cooling. When the cooperative
effect comes into play, the down-ladder process will be accelerated
because the stimulated decay becomes dominant while superradiance
takes place (without other pumping processes such as thermal excitation).
As we have pointed out, the induced rate can be a few orders of magnitude
by increasing the number of particles and hence the cooperativity.
This suggests a scheme of {}``superradiance-assisted cooling''.
Such scheme may be an alternative to cool vibrational states of molecules,
which, generally speaking, has previously been obstructed by the fact
that such states are only weakly optically coupled.

\section*{APPENDIX}

\subsection{Two-body master equation}

\begin{figure}
\begin{centering}
\includegraphics[width=6cm]{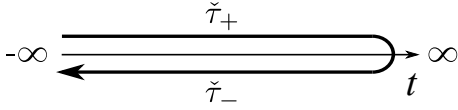}
\par\end{centering}

\caption{\label{fig:contour}Schwinger-Keldysh contour}
\end{figure}

Although this method of the effective two-body description has been
discussed in great details in \cite{PRA99_Fleischhauer,Yelin05},
we summarize in this section the derivation for the formalism for
completeness. We start with the microscopic Hamiltonian that reads
\begin{eqnarray}
H & = & \underbrace{H_{\text{atom}}+H_{\text{field}}-\sum_{j\notin\{1,2\}}\vec{p}_{j}\cdot(\vec{\mathcal{E}}(\vec{r}_{i},t)+\vec{E}(\vec{r}_{j},t))}_{H_{0}}\nonumber \\
 &  & -\underbrace{\sum_{i=1}^{2}\vec{p}_{i}\cdot(\vec{\mathcal{E}}(\vec{r}_{i},t)+\vec{E}(\vec{r}_{i},t))}_{V}.\label{eq:Horigin}
\end{eqnarray}
Here, we separate the field into two parts: the external classical
driving field $\vec{\mathcal{E}}(\vec{r}_{i},t)$ and the induced
local field $\vec{E}(\vec{r}_{i},t)$. When a uniform dense gas of
atoms is considered, it is reasonable to assume that every atom sees
the same field and the same background due to other atoms. On the
other hand, in order to take into account atom-atom quantum correlation
we need to retain adequate degrees of freedom involving at least two
particles. Our proposal is therefore to write down an effective description
for two probe atoms in which all other atoms' contribution will be
averaged in the mean-field sense and appear as parameters in the two-body
description. In Eq. (\ref{eq:Horigin}) $V$ is the interaction of
the probe atoms ($i=1$, $2$) with the field and will be treated
as a small perturbation. $H_{0}$ consists of the unperturbed atomic
and field Hamiltonian as well as the contribution from the background
atoms. In the interaction picture, the evolution operator is then
given by

\begin{equation}
S_{I}(t)=T\exp\Big[-\frac{i}{\hbar}\int_{-\infty}^{t}V_{I}(t^{\prime})dt^{\prime}\Big],\label{eq:si}
\end{equation}
where $T$ is the time-ordering operator. We here introduce the positive
and negative components $x(t)=x^{+}(t)+x^{-}(t)$ with $x^{\pm}(t)=\widetilde{x}(t)e^{\mp i\omega t}$,
where $x_{\mu}\in\{p_{i\mu},E_{\mu},\mathcal{E}_{\mu}\}$ and $\widetilde{x}(t)$
is the slowly varying (compared to the inverse of the radiation frequency
$\omega^{-1}$) amplitude of the corresponding quantity. Further note
that $\widetilde{p}_{i\mu}^{+}=\wp_{\mu}\sigma_{i}^{-}$ and $\widetilde{p}_{i\mu}^{-}=\wp_{\mu}\sigma_{i}^{+}$.
In the rotating wave approximation, the interaction becomes $V_{I}(t)\simeq\sum_{i\mu}[p_{i\mu}^{+}(E_{\mu}^{-}+\mathcal{E}_{\mu}^{-})+\mbox{h.c.}]$.
Eq. (\ref{eq:si}) can be cast in the framework of Schwinger-Keldysh
formalism, in which $V_{I}(t)\longrightarrow V(\check{\tau})$ and
$S_{I}(t)\longrightarrow$
\begin{equation}
S_{C}=T_{C}\exp\Big[-\frac{i}{\hbar}\int_{C}V(\check{\tau})d\check{\tau}\Big],
\end{equation}
where $C$ denotes the Schwinger-Keldysh contour as shown in Fig.
\ref{fig:contour}, and $T_{C}$ is the contour-oriented time-ordering
operator, i.e., along the upper branch of the contour $T_{C}$ is
the normal time-ordering operator while along the lower branch is
the inverse time-ordering operator. To prevent possible confusion
we denote the {}``time'' parameter along the Keldysh contour with
a check sign. We then trace out the degrees of freedom of the fields
and the background atoms, which leads to an effective evolution operator:
\begin{align}
 & S_{C}^{\text{eff}}=\langle S_{C}\rangle_{\text{field}}\nonumber \\
 & =T_{C}\exp\bigg\{\frac{i}{\hbar}\int_{C}d\check{\tau}\sum_{i=1}^{2}\sum_{\mu}\Big[p_{i\mu}^{+}(\check{\tau})\mathcal{E}_{L\mu}^{-}(\vec{r}_{i},\check{\tau})\nonumber \\
 & +p_{i\mu}^{-}(\check{\tau})\mathcal{E}_{L\mu}^{+}(\vec{r}_{i},\check{\tau})\Big]\label{eq:sceff}\\
 & -\frac{1}{2\hbar^{2}}\iint_{C}d\check{\tau}_{1}d\check{\tau}_{2}\sum_{i,j=1}^{2}\sum_{\mu\nu}\Big[p_{i\mu}^{+}(\check{\tau}_{1})D_{i\mu,j\nu}(\check{\tau}_{1},\check{\tau}_{2})p_{j\nu}^{-}(\check{\tau}_{2})\nonumber \\
 & +p_{i\mu}^{-}(\check{\tau}_{1})C_{i\mu,j\nu}(\check{\tau}_{1},\check{\tau}_{2})p_{j\nu}^{+}(\check{\tau}_{2})\Big]\bigg\},\nonumber 
\end{align}
 where $\vec{\mathcal{E}}_{L}^{\pm}(\vec{r}_{i},\check{\tau})=\vec{\mathcal{E}}^{\pm}(\vec{r}_{i},\check{\tau})+\langle\vec{E}^{\pm}(\vec{r}_{i},\check{\tau})\rangle$
is the local field seen by the probe atom, and the Green's function
of the interacting field
\begin{eqnarray}
D_{i\mu,j\nu}(\check{\tau}_{1},\check{\tau}_{2}) & = & \langle\!\langle T_{C}E_{\mu}^{-}(\vec{r}_{i},\check{\tau}_{1})E_{\nu}^{+}(\vec{r}_{j},\check{\tau}_{2})\rangle\!\rangle\\
C_{i\mu,j\nu}(\check{\tau}_{1},\check{\tau}_{2}) & = & \langle\!\langle T_{C}E_{\mu}^{+}(\vec{r}_{i},\check{\tau}_{1})E_{\nu}^{-}(\vec{r}_{j},\check{\tau}_{2})\rangle\!\rangle.
\end{eqnarray}
To get Eq. (\ref{eq:sceff}) we have used

\begin{eqnarray}
\big<T_{C}\exp[s\hat{A}]\big> & = & \exp\bigg[\sum_{m}\frac{s^{m}}{m!}\langle\!\langle T_{C}\hat{A}^{m}\rangle\!\rangle\bigg],
\end{eqnarray}
with $\langle\!\langle\cdot\rangle\!\rangle$ denoting the cumulant,
which is defined by
\begin{eqnarray*}
\langle\!\langle\hat{A}\rangle\!\rangle & = & \langle\hat{A}\rangle\\
\langle\!\langle\hat{A}\hat{B}\rangle\!\rangle & = & \langle\hat{A}\hat{B}\rangle-\langle\hat{A}\rangle\langle\hat{B}\rangle,
\end{eqnarray*}
where $\hat{A}$ and $\hat{B}$ are operators. We also set the higher-order
cumulants $\langle\!\langle E^{m}\rangle\!\rangle=0$ for $m>2$ by
assuming that the radiation field is Gaussian. The two-field Green's
function $D_{i\mu,j\nu}(\check{\tau}_{1},\check{\tau}_{2})$, depending
on the order of $\check{\tau}_{1}$ and $\check{\tau}_{2}$ on $C$,
has four possible forms:
\begin{eqnarray}
D_{\mu\nu}^{++} & = & \langle\!\langle TE_{\mu}^{-}(\vec{r}_{i},\tau_{1+})E_{\nu}^{+}(\vec{r}_{j},\tau_{2+})\rangle\!\rangle\\
D_{\mu\nu}^{--} & = & \langle\!\langle T^{-1}E_{\mu}^{-}(\vec{r}_{i},\tau_{1-})E_{\nu}^{+}(\vec{r}_{j},\tau_{2-})\rangle\!\rangle\\
D_{\mu\nu}^{-+} & = & \langle\!\langle E_{\mu}^{-}(\vec{r}_{i},\tau_{1-})E_{\nu}^{+}(\vec{r}_{j},\tau_{2+})\rangle\!\rangle\\
D_{\mu\nu}^{+-} & = & \langle\!\langle E_{\nu}^{+}(\vec{r}_{j},\tau_{2-})E_{\mu}^{-}(\vec{r}_{i},\tau_{1+})\rangle\!\rangle.
\end{eqnarray}
The other Green's function $C_{i\mu,j\nu}$ has similar relations.
By the subscript {}``+'' or {}``-'' we denote the upper or lower
branch for $\tau$, respectively. In Eq. (\ref{eq:sceff}), those
terms like
\begin{align}
 & \iint_{C}d\check{\tau}_{1}d\check{\tau}_{2}p_{i\mu}^{+}(\check{\tau}_{1})D_{i\mu,j\nu}(\check{\tau}_{1},\check{\tau}_{2})p_{j\nu}^{-}(\check{\tau}_{2})\nonumber \\
 & =\sum_{A,B\in\{+,-\}}\kappa_{AB}\wp_{\mu}\wp_{\nu}\int_{-\infty}^{\infty}d\tau_{1}\int_{-\infty}^{\infty}d\tau_{2}\times\\
 & \qquad\qquad\sigma_{iA}^{-}D_{i\mu,j\nu}^{AB}(\tau_{1},\tau_{2})\sigma_{jB}^{+}e^{i\omega(\tau_{2}-\tau_{1})},\nonumber 
\end{align}
where $\kappa_{AB}=1$ for $A=B$ and $\kappa_{AB}=-1$ for $A\neq B$,
and the subscripts $A,B$ being placed with $\sigma^{-}$ and $\sigma^{+}$
emphasizes that the operators must be in order accordingly on the
Schwinger-Keldysh contour $C$. We change the time variables for the
field correlation such that $D_{i\mu,j\nu}^{AB}(\tau_{1},\tau_{2})\longrightarrow D_{i\mu,j\nu}^{AB}(\tau,\tau^{\prime})$
with $\tau_{1}=\tau-\tau^{\prime}/2$ and $\tau_{2}=\tau+\tau^{\prime}/2$.
After some math, we reach \begin{widetext}
\begin{eqnarray}
S_{C}^{\text{eff}} & = & T_{c}\exp\bigg\{\sum_{i,\mu}\frac{i\wp_{\mu}}{\hbar}\int_{-\infty}^{\infty}d\tau\big[\sigma_{i}^{-}(\tau_{+})\mathcal{E}_{L\mu}^{-}(\vec{r}_{i},\tau)-\sigma_{i}^{-}(\tau_{-})\mathcal{E}_{L\mu}^{-}(\vec{r}_{i},\tau)+\sigma_{i}^{+}(\tau_{+})\mathcal{E}_{L\mu}^{+}(\vec{r}_{i},\tau)-\sigma_{i}^{+}(\tau_{-})\mathcal{E}_{L\mu}^{+}(\vec{r}_{i},\tau)\big]\nonumber \\
 &  & -\int_{-\infty}^{\infty}d\tau\frac{\Gamma_{i\mu,j\nu}(\omega,\tau)}{2}\big[\sigma_{i}^{-}(\tau_{+})\sigma_{j}^{+}(\tau_{+})+\sigma_{i}^{-}(\tau_{-})\sigma_{j}^{+}(\tau_{-})-2\sigma_{i}^{-}(\tau_{-})\sigma_{j}^{+}(\tau_{+})\big]\nonumber \\
 &  & -\int_{-\infty}^{\infty}d\tau\frac{\Gamma_{i\mu,j\nu}(\omega,\tau)+\gamma_{i\mu,j\nu}(\omega,\tau)}{2}\big[\sigma_{j}^{+}(\tau_{+})\sigma_{i}^{-}(\tau_{+})+\sigma_{j}^{+}(\tau_{-})\sigma_{i}^{-}(\tau_{-})-2\sigma_{j}^{+}(\tau_{-})\sigma_{i}^{-}(\tau_{+})\big]\nonumber \\
 &  & +\frac{i}{\hbar}\int_{-\infty}^{\infty}d\tau H_{i\mu,j\nu}(\omega,\tau)\big[\sigma_{i}^{-}(\tau_{+})\sigma_{j}^{+}(\tau_{+})-\sigma_{i}^{-}(\tau_{-})\sigma_{j}^{+}(\tau_{-})-\sigma_{j}^{+}(\tau_{+})\sigma_{i}^{-}(\tau_{+})+\sigma_{j}^{+}(\tau_{-})\sigma_{i}^{-}(\tau_{-})\big]\nonumber \\
 &  & +\frac{i}{\hbar}\int_{-\infty}^{\infty}d\tau h_{i\mu,j\nu}(\omega,\tau)\big[\sigma_{j}^{+}(\tau_{+})\sigma_{i}^{-}(\tau_{+})+\sigma_{j}^{+}(\tau_{-})\sigma_{i}^{-}(\tau_{-})\big]\bigg\},\label{eq:sceffall}
\end{eqnarray}
where we have introduced these quantities:
\begin{eqnarray}
\Gamma_{i\mu,j\nu}(\tau,\omega) & = & \frac{\wp_{\mu}\wp_{\nu}}{\hbar^{2}}\int_{-\infty}^{\infty}d\tau^{\prime}\langle\!\langle E_{\mu}^{-}(\vec{r}_{i},\tau)E_{\nu}^{+}(\vec{r}_{j},\tau+\tau^{\prime})\rangle\!\rangle e^{i\omega\tau^{\prime}},\\
\gamma_{i\mu,j\nu}(\tau,\omega) & = & \frac{\wp_{\mu}\wp_{\nu}}{\hbar^{2}}\int_{-\infty}^{\infty}d\tau^{\prime}\langle\big[E_{\mu}^{+}(\vec{r}_{i},\tau),E_{\nu}^{-}(\vec{r}_{j},\tau+\tau^{\prime})\big]\rangle e^{i\omega\tau^{\prime}},\\
H_{i\mu,j\nu}(\tau,\omega) & = & \frac{i\wp_{\mu}\wp_{\nu}}{2\hbar}\int_{0}^{\infty}d\tau^{\prime}\big\{\langle\!\langle E_{\mu}^{-}(\vec{r}_{i},\tau)E_{\nu}^{+}(\vec{r}_{j},\tau-\tau^{\prime})\rangle\!\rangle e^{-i\omega\tau^{\prime}}-\langle\!\langle E_{\mu}^{-}(\vec{r}_{i},\tau)E_{\nu}^{+}(\vec{r}_{j},\tau+\tau^{\prime})\rangle\!\rangle e^{i\omega\tau^{\prime}}\big\},\label{eq:termH}\\
h_{i\mu,j\nu}(\tau,\omega) & = & \frac{i\wp_{\mu}\wp_{\nu}}{2\hbar}\int_{0}^{\infty}d\tau^{\prime}\big\{\langle\big[E_{\mu}^{-}(\vec{r}_{i},\tau),E_{\nu}^{+}(\vec{r}_{j},\tau-\tau^{\prime})\big]\rangle e^{-i\omega\tau^{\prime}}-\langle\big[E_{\mu}^{-}(\vec{r}_{i},\tau),E_{\nu}^{+}(\vec{r}_{j},\tau+\tau^{\prime})\big]\rangle e^{i\omega\tau^{\prime}}\big\}.\label{eq:termh}
\end{eqnarray}
\end{widetext} From Eq. (\ref{eq:sceffall}) we extract the effective
master equation:
\begin{align}
 & \dot{\rho}(t)=-\frac{i}{\hbar}\big[H_{0},\rho\big]\nonumber \\
 & +\sum_{j=1,2}\sum_{\mu}\frac{i}{\hbar}\wp_{\mu}\big[\sigma_{j}^{-}\mathcal{E}_{L,\mu}^{-}(\vec{r}_{j})+\sigma_{j}^{\dagger}\mathcal{E}_{L,\mu}^{+}(\vec{r}_{j}),\rho\big]\nonumber \\
 & +\frac{i}{\hbar}\sum_{i=1,2}\sum_{\mu,\nu}H_{i\mu,i\nu}\Big[\big[\sigma_{i\mu}^{-},\sigma_{i\nu}^{+}\big],\rho\Big]\label{eq:overall_mastereq}\\
 & -\sum_{i,j=1,2}\sum_{\mu,\nu}\frac{\Gamma_{i\mu,j\nu}}{2}\big(\big[\rho\sigma_{i}^{-},\sigma_{j}^{\dagger}\big]+\big[\sigma_{i}^{-},\sigma_{j}^{\dagger}\rho\big]\big)\nonumber \\
 & -\sum_{i,j=1,2}\sum_{\mu,\nu}\frac{\Gamma_{i\mu,j\nu}+\gamma_{i\mu,j\nu}}{2}\big(\big[\rho\sigma_{j}^{\dagger},\sigma_{i}^{-}\big]+\big[\sigma_{j}^{\dagger},\sigma_{i}^{-}\rho\big]\big).\nonumber 
\end{align}
Note that $\rho$ is the two-body density operator so is a $(2j+1)^{2}\times(2j+1)^{2}$
matrix. Now we identify term $\Gamma$ as the induced pump and decay
rate and term $\gamma$ as the spontaneous decay rate inside the atomic
medium. Term $h$ as shown in Eq. (\ref{eq:termh}) corresponds to
the Lamb shifts, which are somewhat irrelevant for our current consideration
and is therefore absorbed to the unperturbed Hamiltonian $H_{0}$;
term $H$ as shown in Eq. (\ref{eq:termH}) accounts for the collective
light shifts and inhomogeneous broadening. In this paper, we neglect
the dipole shifts and frequency chirping by dropping the diagonal
shift terms, and stress on the quantum correction that makes significance
to the superradiance mechanism. The relevant part of the master equation
now reads

\begin{eqnarray}
\dot{\rho} & = & -\sum_{i,j=1,2}\frac{\Gamma_{ij}}{2}\big(\big[\rho\sigma_{i}^{-},\sigma_{j}^{\dagger}\big]+\big[\sigma_{i}^{-},\sigma_{j}^{\dagger}\rho\big]\big)\nonumber \\
 & - & \sum_{i,j=1,2}\frac{\Gamma_{ij}+\gamma\delta_{ij}}{2}\big(\big[\rho\sigma_{j}^{\dagger},\sigma_{i}^{-}\big]+\big[\sigma_{j}^{\dagger},\sigma_{i}^{-}\rho\big]\big).
\end{eqnarray}
We follow the derivation in \cite{PRA99_Fleischhauer,Yelin05} and
summarize the main results. $\Gamma_{ij}$ can be evaluated through\begin{widetext}
\begin{eqnarray}
\Gamma & = & \frac{\wp^{2}}{\hbar^{2}}\int d^{3}x\left|\widetilde{D}^{ret}(\vec{r}_{0},t|\vec{x},\Delta)\right|^{2}\widetilde{P}^{(1)s}(\vec{r}_{0},t|\Delta)\nonumber \\
 & + & \frac{\wp^{2}}{\hbar^{2}}\int\int d^{3}x_{1}d^{3}x_{2}\widetilde{D}^{ret}(\vec{r}_{0},t|\vec{x}_{1},\Delta)\widetilde{D}^{\ast ret}(\vec{r}_{0},t|\vec{x}_{2},\Delta)\widetilde{P}^{(2)s}(\vec{r}_{0},t|\vec{x}_{1},\vec{x}_{2},\Delta)\label{eq:Gamma1_integration}\\
\bar{\Gamma} & = & \frac{\wp^{2}}{\hbar^{2}}\int d^{3}x\widetilde{D}^{ret}(\vec{r}_{0},t|\vec{x},\Delta)\widetilde{D}^{\ast ret}(\vec{r}_{0},t|\vec{x},\Delta)\widetilde{P}^{(1)s}(\vec{r}_{0},t|\Delta)\nonumber \\
 & + & \frac{\wp^{2}}{\hbar^{2}}\int\int d^{3}x_{1}d^{3}x_{2}\widetilde{D}^{ret}(\vec{r}_{0},t|\vec{x}_{1},\Delta)\widetilde{D}^{\ast ret}(\vec{r}_{0},t|\vec{x}_{2},\Delta)\widetilde{P}^{(2)s}(\vec{r}_{0},t|\vec{x}_{1},\vec{x}_{2},\Delta).\label{eq:Gamma2_integration}
\end{eqnarray}
\end{widetext} We denote $\Gamma\equiv\Gamma_{ii}$ and $\bar{\Gamma}\equiv\Gamma_{ij}$
for $i\neq j$. For above, the retarded function $\widetilde{P}^{ret}$,
the single-particle and two-particle source functions $\widetilde{P}^{(1)s}$
and $\widetilde{P}^{(2)s}$ are given respectively by 
\begin{eqnarray}
\widetilde{P}^{ret}(\vec{r}_{0},t|\Delta) & = & \frac{\mathcal{N}\wp^{2}}{\hbar^{2}}\left[\frac{V(t)}{(\frac{\gamma}{2}+\Gamma)-i\Delta}\right]\label{eq:retarded_P}\\
\widetilde{P}^{(1)s}(\vec{r}_{0},t|\Delta) & = & \frac{\mathcal{N}\wp^{2}}{\hbar^{2}}\left[\frac{2A(t)(\frac{\gamma}{2}+\Gamma)}{(\frac{\gamma}{2}+\Gamma)^{2}+\Delta^{2}}\right]\label{eq:source_P1}\\
\widetilde{P}^{(2)s}(\vec{r}_{0},t|\vec{x}_{1},\vec{x}_{2},\Delta) & = & \frac{\mathcal{N}^{2}\wp^{2}}{\hbar^{2}}\left[\frac{2Y(t)(\frac{\gamma}{2}+\Gamma)}{(\frac{\gamma}{2}+\Gamma)^{2}+\Delta^{2}}\right]\label{eq:source_P2}
\end{eqnarray}
with $\mathcal{N}$ is the particle volume density, and $-\Delta$
accounts for the frequency of the Fourier component relative to that
of the real field. Quantities $A(t)$, $V(t)$, are $Y(t)$ are determined
by the density matrix through Eqs. (\ref{eq:At},\ref{eq:Vt},\ref{eq:Yt}).
In addition, The retarded Green's function is 
\begin{equation}
\widetilde{D}^{ret}(\vec{r}_{0},t|\vec{x},\Delta)=-\frac{i\hbar\omega^{2}}{6\pi\epsilon_{0}c^{2}}\frac{e^{q_{0}^{\prime\prime}x}}{x}e^{-iq_{0}^{\prime}x}\label{eq:retarded_D}
\end{equation}
with $q_{0}^{\prime}=\omega/c$ and $q_{0}^{\prime\prime}=\frac{\hbar\omega}{3\epsilon_{0}c}\widetilde{P}^{ret}(\Delta)$.
Through direct integration for Eqs. (\ref{eq:Gamma1_integration},\ref{eq:Gamma2_integration}),
we finally get
\begin{eqnarray}
\Gamma & = & \gamma(e^{2\zeta}-1)\frac{A(t)}{V(t)}+2\mathcal{C}^{2}\varrho^{4}\frac{\gamma^{2}I(\zeta,\tilde{\varrho})}{\Gamma+\gamma/2}Y(t)\label{eq:Gamma1-1}\\
\bar{\Gamma} & = & \frac{\gamma^{2}I(\zeta,\tilde{\varrho})}{\Gamma+\gamma/2}\left[3\mathcal{C}\varrho A(t)+2\mathcal{C}^{2}\varrho^{4}Y(t)\right],\label{eq:Gamma2-1}
\end{eqnarray}
with $\zeta(\Delta)\equiv\frac{1}{2}\mathcal{C}\varrho\frac{\gamma(\Gamma+\gamma/2)}{(\Gamma+\gamma/2)^{2}+\Delta^{2}}V(t)$,
$\tilde{\varrho}(\Delta)\equiv\omega d/(2c)-\frac{\Delta}{(\Gamma+\gamma/2)}\zeta(\Delta)$,
and $I(\zeta,\tilde{\varrho})\equiv[\big((\zeta-1)e^{\zeta}+\cos\tilde{\varrho}\big)^{2}+\big(\tilde{\varrho}e^{\zeta}-\sin\tilde{\varrho}\big)^{2}]/(\zeta^{2}+\tilde{\varrho}^{2})^{2}$.
For large $\tilde{\varrho}$ and $\zeta$, $I(\tilde{\varrho})$ can
be approximated by $e^{2\zeta}/(\zeta^{2}+\tilde{\varrho}^{2})$.
We set $\Delta=0$ for the resonant case.

Note that $\zeta=\zeta(\mathcal{C}\varrho)$ and the second term in
Eq. (\ref{eq:Gamma1-1}) is proportional to $\mathcal{C}^{2}\varrho^{4}e^{2\zeta}/(\zeta^{2}+\varrho^{2})\sim\mathcal{C}^{2}\varrho^{2}e^{2\zeta}$
when $\varrho\gg\zeta$. This indicates that the superradiant dynamics
is, roughly speaking, characterized by the most relevant parameter
$\mathcal{C}\varrho$ although the dependence on $\mathcal{C}$ and
$\varrho$ individually can be evident, but less significant, through
the exact form of Eqs. (\ref{eq:Gamma1-1}, \ref{eq:Gamma2-1}).
\begin{acknowledgments}
The authors wish to thank Marc Repp, Juris Ulmanis, Johannes Deiglmayr,
and Matthias Weidemüller for helpful input. We thank the NSF for financial
support.\end{acknowledgments}


\begin{thebibliography}{References}
\bibitem{Sci95_Anderson} M. H. Anderson, J. R. Ensher, M. R. Matthews,
C. E. Wieman, and E. A. Cornell, \textit{Science} \textbf{269}, 198
(1995). 

\bibitem{Nat03_Greiner} M. Greiner, C. A. Regal, and D. S. Jin, \textit{Nature}
\textbf{426}, 537 (2003). 

\bibitem{RMP01_Leggett} A. J. Leggett, Rev. Mod. Phys. \textbf{73},
307 (2001).

\bibitem{PRB89_Fisher} M. P. A. Fisher, P. B. Weichman, G. Grinstein,
and D. S. Fisher, \textit{Phys. Rev.} B \textbf{40}, 546 (1989).

\bibitem{Nat02_Greiner} M. Greiner, O. Mandel, T. Esslinger, T. W.
Hansch, and I. Bloch, \textit{Nature} \textbf{415}, 39 (2002).

\bibitem{Nat06_Hadzibabic} Z. Hadzibabic, P. Kruger, M. Cheneau,
B. Battelier, and J. Dalibard, \textit{Nature} \textbf{441}, 1118
(2006). 

\bibitem{PRL08_Weimer} H. Weimer, R. Löw, T. Pfau, and H. P. Büchler,
\textit{Phys. Rev. Lett.} \textbf{101}, 250601 (2008). 

\bibitem{PRL03_Duan} L.-M. Duan, E. Demler, and M. D. Lukin, \textit{Phys.
Rev. Lett.} \textbf{91}, 090402 (2003). 

\bibitem{PRL04_Porras} D. Porras and J. I. Cirac, \textit{Phys. Rev.
Lett.} \textbf{92}, 207901 (2004).

\bibitem{NatPhys08_Friedenauer} A. Friedenauer, H. Schmitz, J. T.
Glueckert, D. Porras, and T. Schaetz, \textit{Nature Physics} \textbf{4},
757 (2008).

\bibitem{Nat10_Gerritsma} R. Gerritsma, G. Kirchmair, F. Zahringer,
E. Solano, R. Blatt, and C. F. Roos, \textit{Nature} \textbf{463},
68 (2010). 

\bibitem{Nat10_Kim} K. Kim, M.-S. Chang, S. Korenblit, R. Islam,
E. E. Edwards, J. K. Freericks, G.-D. Lin, L.-M. Duan, and C. Monroe,
\textit{Nature} \textbf{465}, 590 (2010).

\bibitem{Nat09_Lin} Y.-J. Lin, R. L. Compton, K. Jimenez-Garcia,
J. V. Porto, and I. B. Spielman, \textit{Nature} \textbf{462}, 628
(2009).

\bibitem{PRL99_Brennen} G. K. Brennen, C. M. Caves, P. S. Jessen,
and I. H. Deutsch, \textit{Phys. Rev. Lett.} \textbf{82}, 1060 (1999).

\bibitem{Kuznetsova11a} E. Kuznetsova, T. Bragdon, R. Côté, and S.
F. Yelin, arXiv:quant-ph/1106.0713v2 (2011). 

\bibitem{Kuznetsova11b} E. Kuznetsova, S. T. Rittenhouse, H. R. Sadeghpour,
and S. F. Yelin, arXiv:quant-ph/1105.2010v1 (2011).

\bibitem{PhysRev54_Dicke} R. H. Dicke, \textit{Phys. Rev.} \textbf{93},
99 (1954).

\bibitem{PhysLett74_DeMartini} F. De Martini and G. Preparata, \textit{Physics
Letters} A \textbf{48}, 43 (1974).

\bibitem{PhysRep82_Gross} M. Gross and S. Haroche, \textit{Physics
Reports} \textbf{93}, 301 (1982).

\bibitem{Sci99_Inouye} S. Inouye, A. P. Chikkatur, D. M. Stamper-Kurn,
J. Stenger, D. E. Pritchard, and W. Ketterle, \textit{Science} \textbf{285},
571 (1999).

\bibitem{PRL99_Moore} M. G. Moore and P. Meystre, \textit{Phys. Rev.
Lett.} \textbf{83}, 5202 (1999).

\bibitem{PRA00_Mustecaplioglu} O. E. Müstecaplioglu and L. You, \textit{Phys.
Rev.} A \textbf{62}, 063615 (2000).

\bibitem{PRL03_Farooqi} S. M. Farooqi, D. Tong, S. Krishnan, J. Stanojevic,
Y. P. Zhang, J. R. Ensher, A. S. Estrin, C. Boisseau, R. Côté, E.
E. Eyler, and P. L. Gould, \textit{Phys. Rev. Lett.} \textbf{91},
183002 (2003).

\bibitem{PRL08_Akkermans} E. Akkermans, A. Gero, and R. Kaiser, \textit{Phys.
Rev. Lett.} \textbf{101}, 103602 (2008).

\bibitem{Nat10_Baumann} K. Baumann, C. Guerlin, F. Brennecke, and
T. Esslinger, \textit{Nature} \textbf{464}, 1301 (2010).

\bibitem{PRL10_Nagy} D. Nagy, G. Kónya, G. Szirmai, and P. Domokos,
\textit{Phys. Rev. Lett.} \textbf{104}, 130401 (2010).

\bibitem{PRA10_Meiser} D. Meiser and M. J. Holland, \textit{Phys.
Rev.} A \textbf{81}, 033847; \textit{ibid.}, 063827 (2010).

\bibitem{PRA07_Wang} T.Wang, S. F. Yelin, R. Côté, E. E. Eyler, S.
M. Farooqi, P. L. Gould, M. Koštrun, D. Tong, and D. Vrinceanu, \textit{Phys.
Rev.} A \textbf{75}, 033802 (2007).

\bibitem{NatPhys07_Scheibner} M. Scheibner, T. Schmidt, L. Worschech,
A. Forchel, G. Bacher, T. Passow, and D. Hommel, \textit{Nature Physics}
\textbf{3}, 106 (2007).

\bibitem{PRA99_Fleischhauer} M. Fleischhauer and S. F. Yelin, \textit{Phys.
Rev.} A \textbf{59}, 2427 (1999).

\bibitem{Yelin05} S. F. Yelin, M. Kostrun, T. Wang, and M. Fleischhauer,
arXiv:quant-ph/0509184 (2005). (To appear in a similar form in \textit{Advances
in Atomic, Molecular, and Optical Physics} Vol. \textbf{61}).

\bibitem{Deiglmayr11} J. Deiglmayr, M. Repp, O. Dulieu, R. Wester,
and M. Weidemüller, \textit{Eur. Phys. J.} D, in press.\end{thebibliography}
\end{document}